\documentclass[pra,aps,twocolumn,nopacs,superscriptaddress,nofootinbib]{revtex4-1}

\usepackage{amsmath}  \usepackage{amssymb}  \usepackage{amsfonts}  \usepackage{bm}  \usepackage{bbm}   \usepackage{braket}   \usepackage{comment}  \usepackage{dcolumn}  \usepackage{enumerate}  \usepackage{epsfig}  \usepackage{gensymb}  \usepackage{graphicx}  \usepackage{indentfirst}  \usepackage{mathtools}  \usepackage{psfrag}  \usepackage{pst-all}  \usepackage{soul}  \usepackage{siunitx}   \usepackage{xcolor}  \usepackage{microtype} \usepackage{txfonts} \usepackage{txfontsb} 
\usepackage{float} 
\usepackage[colorlinks,linkcolor=blue,citecolor=blue,urlcolor=blue]{hyperref} 
\usepackage[T1]{fontenc} 

\def\ii{{\rm i}}  \def\ee{{\rm e}}
\def\me{m_{\rm e}}  
  
\def\Ab{{\bf A}}        \def\Eb{{\bf E}}            \def\Hb{{\bf H}}                      \def\qb{{\bf q}}    \def\rb{{\bf r}}      \def\vb{{\bf v}} 
\def\xx{\hat{\bf x}}  \def\yy{\hat{\bf y}}  \def\zz{\hat{\bf z}}            

      \def\Hint{{\mathcal{H}}^{\rm int}}  \def\wmin{{\omega_{\rm min}}}  \def\wmax{{\omega_{\rm max}}}    \def\wt{\tilde{\omega}}  \def\Dt{\tilde{\Delta}}  \def\sigmat{\sigma}
\def\zt{z-vt}  \def\ztp{(z-vt)}

\begin{document} 
\def\bibsection{\section*{\refname}}

\title{Zeptosecond free-electron compression through temporal lensing}

\author{Xin~Jin}
\affiliation{ICFO--Institut de Ciencies Fotoniques, The Barcelona Institute of Science and Technology, 08860 Castelldefels (Barcelona), Spain}
\author{Cruz~I.~Velasco}
\affiliation{ICFO--Institut de Ciencies Fotoniques, The Barcelona Institute of Science and Technology, 08860 Castelldefels (Barcelona), Spain}
\author{F.~Javier~Garc\'{\i}a~de~Abajo}
\email[E-mail: ]{javier.garciadeabajo@nanophotonics.es}
\affiliation{ICFO--Institut de Ciencies Fotoniques, The Barcelona Institute of Science and Technology, 08860 Castelldefels (Barcelona), Spain}
\affiliation{ICREA--Instituci\'o Catalana de Recerca i Estudis Avan\c{c}ats, Passeig Llu\'{\i}s Companys 23, 08010 Barcelona, Spain}

\begin{abstract}
The pursuit of ever-shorter time scales is a frontier in modern physics, exemplified by the synthesis of attosecond light pulses---an achievement made possible by coherently superimposing a broad range of photon energies, as required by the uncertainty principle. However, extending this progress into the zeptosecond regime poses significant challenges, as it demands phase-correlated optical spectra spanning hundreds of electronvolts. In this context, electrons offer a compelling alternative to light because they can be coherently manipulated to form broad energy superpositions, as demonstrated by the generation of attosecond pulses in ultrafast electron microscopes. Here, we propose a practical scheme for compressing free electrons into the zeptosecond domain by modulating their wave functions using suitably tailored broadband light fields. Building on recent advances in {free-electron--light--matter} interactions, our method introduces the concept of temporal lensing---an extension of conventional optical lensing to the time domain---to produce electron pulses with arbitrarily short durations.
\end{abstract}

\maketitle
\date{\today}

\section{Introduction}

Attosecond science is enabled by the synthesis of short light pulses resulting from the superposition of optical high-order harmonics, which are commonly generated by irradiating atomic gases with intense infrared laser pulses \cite{CK07,KI09,CMC22,A24}. In typical experiments, a separate component of each driving pulse is used to pump a specimen and trigger ultrafast nonlinear dynamics, while the generated X-ray attosecond pulse probes the system after a controlled delay time. The scattered X-rays yield time-resolved information on the evolution of different physical quantities, such as the intra-atomic dynamics following photoemission from a core level \cite{DHK02,NEB12} or the energy modulation of bound electrons in a conductive material driven by the pumping near field \cite{IQY02,SKK07}. Photoelectron rescattering induced by intense-field irradiation of a free-space molecule has also been used to monitor attosecond-scale dynamics within the ionized target \cite{WPL16}. Pushing the duration of optical pulses below 1~as requires the superposition of harmonics extending beyond $\sim330$~eV, as required by the uncertainty principle in the optimum scenario that their amplitudes are shaped in a Gaussian-like spectral profile. Consequently, reaching the zeptosecond regime is challenging \cite{HPP13} and currently unattainable with light.

Free electrons emerge as an appealing alternative for producing short probes. In the so-called photon-induced near-field electron microscopy (PINEM) technique \cite{BFZ09,paper151,FES15,PRY17,B17_2,MB18_2,SMY19,RTN20,DNS20,paper371}, the electron wave function is spectrally shaped by interaction with material-scattered optical fields, which generate a periodic energy comb corresponding to the stimulated absorption or emission of different numbers of photons. Similar to the combination of optical harmonics to produce X-ray pulses, electron energy sidebands travelling at different velocities are eventually superimposed after free-space propagation of the electron, giving rise to trains of ultrashort pulses in the attosecond regime \cite{PRY17,B17_2,MB18_2,SMY19,RTN20}. Experiments have demonstrated the generation of over one thousand sidebands spanning an energy interval in the keV range \cite{DNS20}, and even broader spectra are possible by increasing the employed light intensity \cite{paper371}. Recently, subcycle optical-near-field dynamics in the attosecond regime has been measured using homodyne electron detection \cite{NKS23,paper431,BNS24}. Beyond near-field approaches, free-space inelastic electron--light scattering (IELS) has also been shown to produce electron energy sidebands \cite{KES18,KSH18,TTB23} and has been proposed as an effective method for temporally modulating the electron wave function \cite{paper445}. In this context, both near-field \cite{paper351,paper397} and ponderomotive \cite{paper368,MWS22} IELS have been demonstrated to shape the transverse electron wave function.

\begin{figure*}[ht!]
\centering\includegraphics[width=1.0\textwidth]{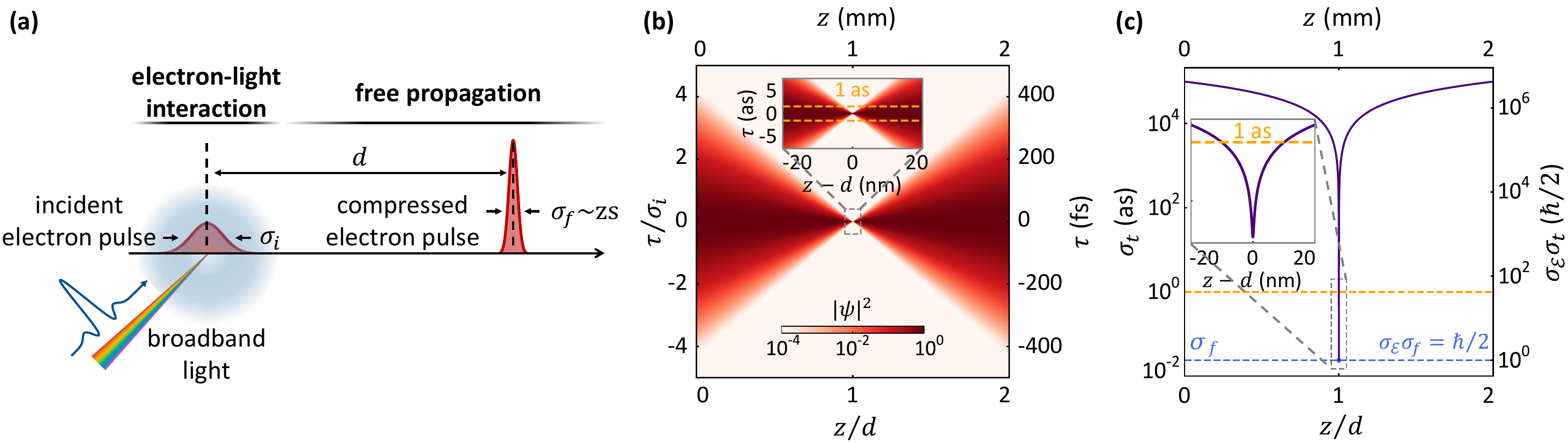}
\caption{\textbf{Free-electron compression via temporal lensing.}
\textbf{(a)}~Schematic of an incident Gaussian electron wave packet (left) of temporal standard deviation $\sigma_i\sim100$~fs interacting with a broadband evanescent field. After free-space propagation over a distance $d$ beyond the interaction region, the electron is compressed and features a final standard deviation $\sigma_f<1$~as.
\textbf{(b)}~Evolution of the electron probability density $|\psi(z,t)|^2$ as a function of shifted time $\tau=t-z/v$ (vertical axis) and propagation distance $z$ (horizontal axis) starting from the interaction region $z=0$ and passing through the focal position $z=d$. The density is normalized to the maximum value for each $z$.
\textbf{(c)}~Corresponding evolution of the temporal standard deviation $\sigma_t$ (left scale) and the product of energy and time uncertainties $\sigma_\mathcal{E}\sigma_t$ (right scale, in units of $\hbar/2$). Panels (b,c) are calculated for a 200~keV electron (velocity $v\approx0.7\,c$) and $d=1$~mm.}
\label{Fig1}
\end{figure*}

Here, we propose an approach to generate single zeptosecond electron pulses by shaping the electron wave function through the interaction with suitably designed broadband electromagnetic fields. A practical example is provided in which 100~fs single-electron wave packets with 200~keV energy, such as those generated by photoemission with ultrashort laser pulses, are compressed to zeptoseconds after interaction with a GHz field, followed by free propagation over a millimetric distance $d$ (Fig.~\ref{Fig1}a). Remarkably, the proposed approach is tolerant to temporal uncertainty in the arrival time of incident electrons. We theoretically demonstrate the feasibility of this method and formulate general principles for its application in the generation of arbitrarily short electron pulses.

\section{Results}

We rely on IELS between a free electron and a shaped optical pulse to imprint a specific spectral profile onto the electron wave function as prescribed by the inverse time-propagation of the desired compressed electron. Unlike PINEM, where quasimonochromatic light pulses transform the electron into a temporal pulse train, our method prepares the electron to evolve into a single compressed pulse after free propagation over a designated distance $d$ from the interaction region (Fig.~\ref{Fig1}b). The standard deviations of the electron pulse in energy and time ($\sigma_\mathcal{E}$ and $\sigma_t$) are subject to the uncertainty principle ($\sigma_\mathcal{E}\sigma_t\ge\hbar/2$), which is saturated (minimum product) at the point of maximum compression when applying optimum illumination. Although $\sigma_\mathcal{E}$ does not change during free propagation, $\sigma_t$ shrinks until a dip in $\sigma_\mathcal{E}\sigma_t$ is produced at the compression region (Fig.~\ref{Fig1}c).

\subsection{Optimal electron compression by interaction with broadband light}

Upon interaction with light, the incident electron wave function $\psi^{\rm inc}(z-vt)$ is modified through a spatiotemporally dependent phase factor $\ee^{\ii\varphi(z-vt)}$, where
\begin{subequations} \label{varphibetaw}
\begin{align}  
\varphi(z)=\ii\int\frac{d\omega}{2\pi}\;\beta_\omega\;\ee^{\ii\omega z/v} \label{phibeta}
\end{align}
admits the spectral decomposition
\begin{align}  
\beta_\omega=\frac{e}{\hbar\omega}\int dz\;E_{z,\omega}(z)\;\ee^{-\ii\omega z/v} \label{betaw}
\end{align}
\end{subequations}
in terms of the frequency-resolved optical electric field $E_{z,\omega}(z)$ along the electron trajectory $z=vt$ (see Methods). Here, we assume a constant electron velocity $v$ (nonrecoil approximation) along with a collimated and narrow electron beam. Under these conditions, the optical field can be considered approximately uniform in the transverse directions, and we can neglect transverse electron motion. Free propagation after IELS is then described by a modified wave function
\begin{align} \label{phif} 
\psi(z,t)\propto\int dz'\;
\psi^{\rm inc}(z')\;\ee^{\ii\varphi(z')+\ii\pi(\zt-z')^2/vt\lambda_e},
\end{align}
where $\lambda_e=2\pi\hbar/(\me v\gamma^3)$ is the de Broglie wavelength of the electron divided by $\gamma^2=1/(1-v^2/c^2)$ (see Methods). Based on this expression, the temporal evolution of an incident Gaussian electron wave packet is illustrated in Fig.~\ref{Fig1}b,c for an initial wave packet duration $\sigma_i$. The wave packet reaches a minimum temporal standard deviation $\sigma_f$ at the designated propagation distance $z=d$ (Fig.~\ref{Fig1}b). For $\sigma_i=100$~fs and $d=1$~mm, we find that 200~keV electrons can be compressed down to $\sigma_f<1$~as, provided an optimum phase profile is introduced (see below). The minimum in temporal standard deviation at $z=d$ coincides with the condition of saturation of the uncertainty relation $\sigma_\mathcal{E}\sigma_t=\hbar/2$ (Fig.~\ref{Fig1}c).

\begin{figure*}[ht!]
\centering\includegraphics[width=1.0\textwidth]{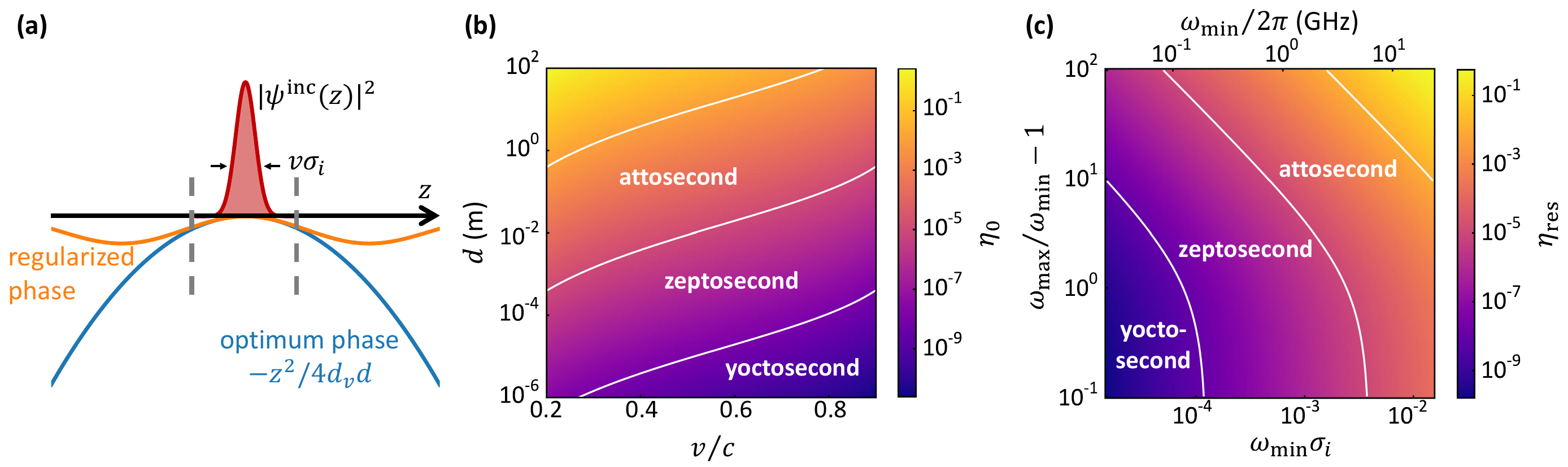}
\caption{\textbf{Optimal temporal compression domains.}
\textbf{(a)}~Phase imprinted onto the electron for optimal compression (blue curve), and regularized phase achieved with a practical illumination scheme avoiding the large-$z$ divergence (orange curve) as a function of position along the incident electron density profile (shaded curve).
\textbf{(b)}~Optimal compression ratio $\eta_0=\sigma_f/\sigma_i$ as a function of electron velocity $v$ and focal distance $d$.
\textbf{(c)}~Residual compression ratio $\eta_{\rm res}$ plotted as a function of the scaled minimum frequency $\wmin\sigma_i$ and the fractional spectral range $\wmax/\wmin-1$ as obtained using illumination with a finite frequency range. The labels in (b,c) and the upper scale in (c) correspond to an initial electron pulse duration $\sigma_i=100$~fs.}
\label{Fig2}
\end{figure*}

Starting with an incident electron wave packet characterized by a temporal standard deviation $\sigma_i$, an arbitrary degree of compression down to $\sigma_f\ll\sigma_i$ is possible for a suitable choice of the imprinted phase profile $\varphi(z)$. More precisely, for compression at a distance $d$ from the IELS region (Fig.~\ref{Fig1}a), we obtain a compression ratio
\begin{subequations}
\label{sseta}
\begin{align} \label{ratio3} 
\frac{\sigmat_f}{\sigmat_i}=\sqrt{\eta_0^2+\eta_{\rm res}^2}
\end{align}
where
\begin{align} \label{ratio2} 
\eta_0=\frac{\lambda_e d}{4\pi v^2\sigmat_i^2}
\end{align}
\end{subequations}
is a lower bound determined by the chosen values of $d$, $v$, and $\sigmat_i$, while $\eta_{\rm res}$ is a residue depending on the optical field (see Methods). In particular, an optimum ratio $\sigma_f/\sigma_i=\eta_0$ (i.e., $\eta_{\rm res}=0$) is achieved with an imprinted phase
\begin{align} \label{varphioptimum} 
\varphi(z)=-\frac{\pi z^2}{\lambda_e d}
\end{align}
(see Methods). The optimum level of compression limited by $\eta_0$ is illustrated in Fig.~\ref{Fig2}b as a function of electron velocity $v$ and propagation distance $d$ for an initial wave packet duration $\sigma_i=100$~fs. At a typical electron-microscope energy of $200$~keV, reaching the zeptosecond regime requires propagation distances of at most a few millimeters.

\begin{figure*}[ht!] 
\centering\includegraphics[width=0.7\textwidth]{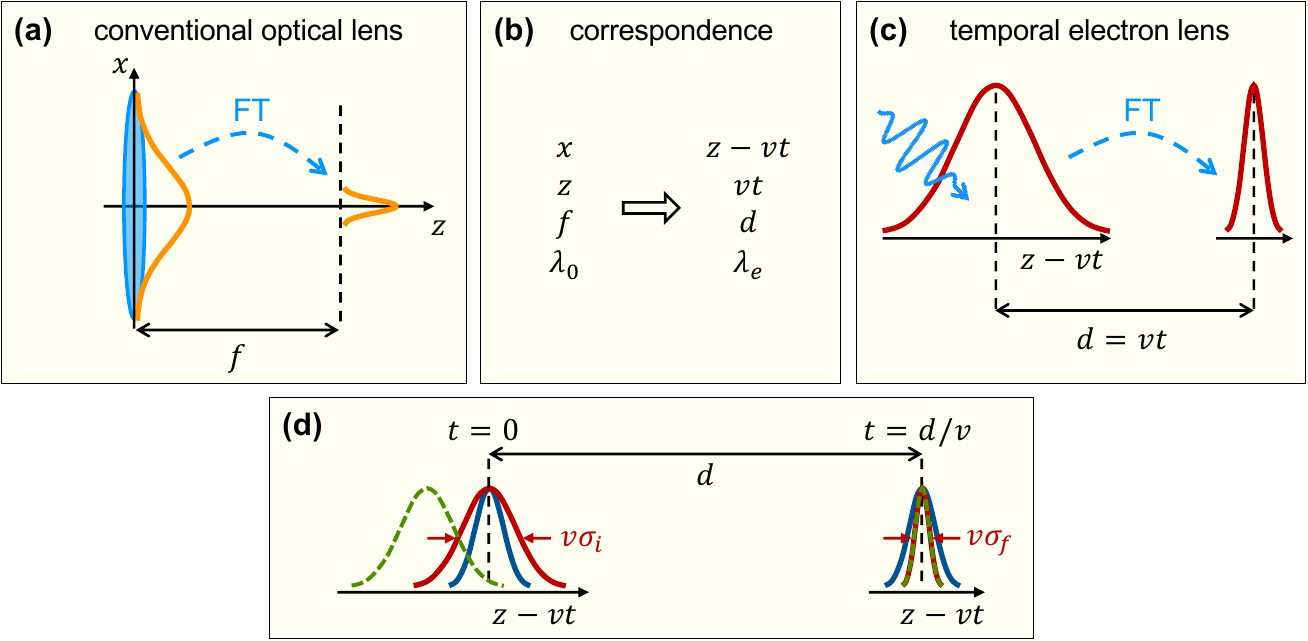}
\caption{\textbf{Conventional optical lensing vs temporal electron lensing.}
\textbf{(a)}~The optical image formed at the focal plane of a lens from an object described by an amplitude function placed at the lens plane is the Fourier transform (FT) of that function. \textbf{(b,c)}~Following the coordinate correspondence in (b), with the optical wavelength $\lambda_0$ replaced by an effective electron wavelength $\lambda_e$, we introduce external illumination acting as a temporal lens on an incident electron wave function, which is similarly imaged as its FT at a focal distance $d$ (c) (i.e., a time $t=d/v$). \textbf{(d)}~Like in conventional optics, the product of the temporal widths of the {\it object} and the {\it image} (i.e., $\sigma_i\sigma_f$) remains constant. Importantly, any temporal shift in the incident electron wave packet---such as that caused by jitter in the electron emission---does not affect the image intensity, effectively {\it erasing} the shift.}
\label{Fig3}
\end{figure*}

The $z^2$ dependence of the imprinted phase in Eq.~(\ref{varphioptimum}) is a reminiscence of the quadratic lateral profile introduced by optical lenses in the paraxial approximation, therefore indicating that we are dealing with a form of focusing, but in the temporal domain, with short free-propagation distances $d$ and large electron velocities $v$ favoring compression. The analogy with conventional optics is further elaborated in Fig.~\ref{Fig3}. Temporal electron lensing as described by Eq.~(\ref{phif}) is illustrated in Fig.~\ref{Fig3}c. At the focus ($vt=d$), the wave function becomes the Fourier transform of the incident one. With the coordinate and wavelength transformation summarized in Fig.~\ref{Fig3}b, this process is shown to be completely analogous to the formation of an image at the focal plane of a conventional optical lens when the object is placed at the lens plane (Fig.~\ref{Fig3}a). Consequently, in both scenarios, the product of the image and object widths (i.e., $\sigma_i\sigma_f=\lambda_ed/4\pi v^2$ for the electron) remains constant, as predicted by Eqs.~(\ref{sseta}) for perfect lensing (i.e., with $\eta_{\rm res}=0$). Note that this concept relies on the wave nature of electrons and is conceptually different from the proposal of using ponderomotive forces to realize ray-like temporal lensing of point-particle electrons \cite{BZ07,HUB09}. In our scheme, the object--image relation is inherited from the wave nature of the electrons and the ensuing properties of the Fourier transform (Fig.~\ref{Fig3}d): besides the constancy of $\sigma_i\sigma_f$, we find that both the image intensity and its temporal position are unaffected by any temporal shift in the incident electron wave packet, such as those caused by uncontrolled jitter in the electron source. Temporal imaging is therefore robust against temporal misalignment between the incident electron pulse and the shaped optical field.

\subsection{Practical realization of zeptosecond electron compression}

Although the phase prescribed by Eq.~(\ref{varphioptimum}) exhibits a pathological divergence with increasing $z$, the incident electron (Fig.~\ref{Fig2}a) only needs to be modified within a finite $z$ region. Therefore, we seek a practical illumination scheme that approximates the profile in Eq.~(\ref{varphioptimum}) over the spatial extension of the incident electron wave packet. After examining different possibilities (see Appendix~\ref{appendixD}), we find it convenient to constrain the illumination to a finite frequency range $\wmin<\omega<\wmax$, resulting in an optical pulse that interacts with the incident electron wave packet within a small spatial region compared to the optical wavelength. To prevent material damage in the electron--light coupling structure (see below), we additionally minimize the optical pulse energy $\mathcal{E}_{\rm opt}$. Maximum compression---corresponding to a minimum residual fraction $\eta_{\rm res}$---is obtained when the coupling coefficient $\beta_\omega$ is frequency-independent (see Methods). For an incident electron Gaussian wave packet, $\eta_{\rm res}$ depends solely on the dimensionless parameters $\wmin\sigma_i$ and $\wmax\sigma_i$ and is plotted in Fig.~\ref{Fig2}c. These results define the achievable compression regimes, illustrated by the solid curves for a representative Gaussian duration of the incident electron $\sigma_i=100$~fs, which requires illumination by GHz-frequency fields to reach the zeptosecond range.

\begin{figure*}[ht!] 
\centering\includegraphics[width=1.0\textwidth]{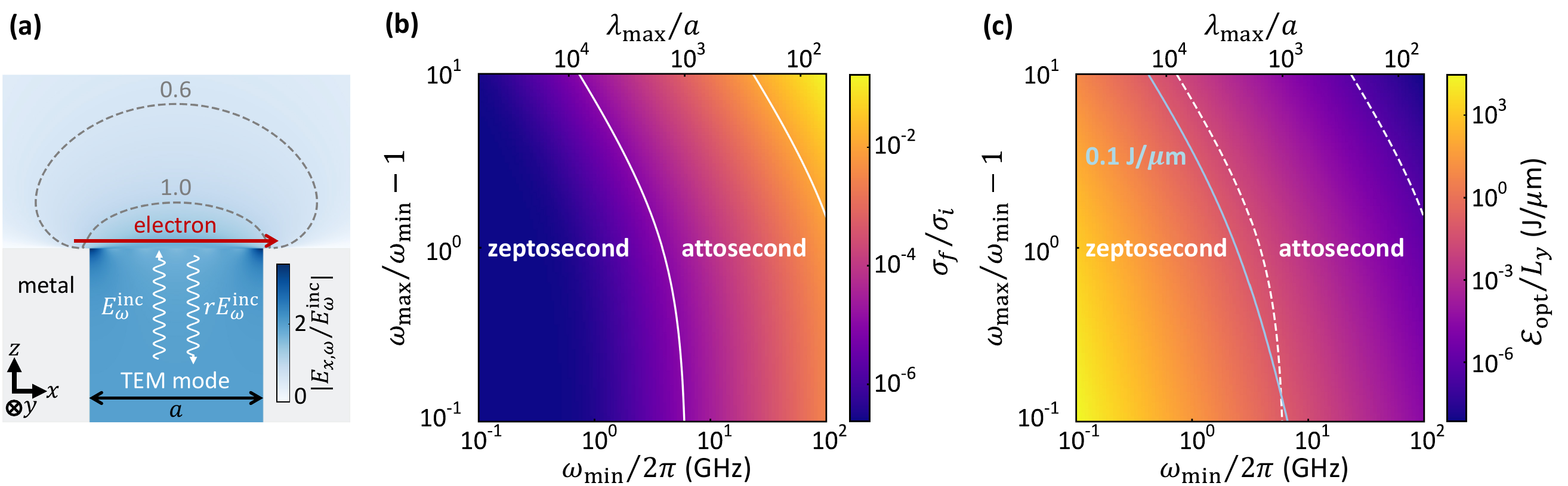}
\caption{\textbf{Zeptosecond electron compression by exposure to waveguided GHz fields.}
\textbf{(a)}~Schematic of the proposed configuration for electron--field interaction, consisting of a slit waveguide of width $a$ flanked by a metal, where TEM modes propagate to the aperture (incident electric field amplitude $E_\omega^{\rm inc}$), bounce back (reflection coefficient $r\approx1$), and spill out to the external vacuum region traversed by the electron. The density plot and dashed contours show the electric field amplitude along the direction of the electron velocity $\vb\parallel\xx$ for a wavelength-to-width ratio of $5000$.
\textbf{(b)}~Compression ratio $\sigma_f/\sigma_i$ obtained with an optimized field spanning a finite spectral range as a function of minimum frequency and fractional spectral range for 200~keV electrons, $\sigma_i=100$~fs, $a=50\,\mu$m, and $d=1$~mm. The upper horizontal scale shows the maximum wavelength $\lambda_{\rm max}=2\pi c/\omega_{\rm min}$ normalized to the slit width.
\textbf{(c)}~Optical energy $\mathcal{E}_{\rm opt}$ carried by optimized pulses under the conditions of (b), normalized to the length of the slit $L_y$ along $y$.}
\label{Fig4}
\end{figure*}

For simplicity, we assume the IELS region to be sufficiently small ($\ll d$) to avoid any significant reshaping of the electron wave function, as would occur during free propagation over a distance $d$ in the millimeter range. We thus need to expose the electron to GHz fields (centimeter wavelength) confined within micron-sized regions, such as those emanating from a metallic slit waveguide perforated in a planar surface (Fig.~\ref{Fig4}a). The confinement region is determined by the slit width $a$. Such a waveguide can transmit transverse electromagnetic (TEM) modes with a high level of spatial compression \cite{T01,SPK09}. For an electron passing close to the slit aperture, our analysis (see Methods and Appendix~\ref{appendixF}) shows that the coupling coefficient is given by $\beta_\omega\approx(2ea/\hbar\omega)\,E^{\rm inc}_\omega$, where $E^{\rm inc}_\omega$ is the frequency-resolved incident TEM mode amplitude inside the waveguide.

For a practical realization of this idea, we consider a waveguide of width $a=50\,\mu$m (much larger than the metal skin depth $\sim1\,\mu$m at GHz frequencies). For a 200~keV electron with an initial duration $\sigma_i=100$~fs and a free propagation distance $d=1$~mm, the final electron pulse duration is plotted in Fig.~\ref{Fig4}b, showing that the zeptosecond regime is reached within a wide range of $\wmax/\wmin$ ratios for $\wmin$ in the GHz and sub-GHz region. The incident optical energy $\mathcal{E}_{\rm opt}$ carried per pulse is represented in Fig.~\ref{Fig4}c as calculated from the Poynting vector (see Methods). The zeptosecond region identified in Fig.~\ref{Fig4}b requires an energy $\sim1$~J per micron of length along $y$ (i.e., the transverse waveguide direction, see Fig.~\ref{Fig4}a). Incidentally, optical heating of the material can compromise its integrity. A detailed analysis (see Methods and Appendix~\ref{appendixF}) yields an upper temperature bound that remains well below the melting temperature of tungsten across a broad range of the attosecond regime shown in Fig.~\ref{Fig4}c.

\section{Concluding remarks}

The proposed scheme for temporal compression is based on the analogy with conventional optical lensing, developed here for objects placed at a lens and imaged at its focal plane. However, more general configurations can be envisioned---for instance, using concatenated temporal lenses to achieve increasingly higher levels of electron compression. Each stage in such a multi-step setup would operate over a distinct spectral range, constrained by the condition $\wmin\sigma_i\ll1$. Unlike the single-step approach, individual steps could implement more moderate compression ratios $\sigma_f/\sigma_i$, thereby relaxing the requirement for extremely low values of $\wmin$. Additionally, this multi-step configuration could alleviate optical heating constraints because the optical energy would be more evenly distributed across the system. In a complementary direction, the established analogy suggests that strategies developed for aberration correction in optical systems could be leveraged to enhance electron temporal compression and pulse shaping.

Compression could also be achieved through illumination over a larger region where electron recoil occurs, allowing for a continuous interplay between electron wave-function reshaping and light-assisted lensing. This approach would further relax the constraints on minimizing optical energy and avoiding material damage, since the illumination would be distributed over a larger spatial area. Implementing this strategy would require a straightforward extension of the present theory.

Although we have considered compression of initial Gaussian pulses for simplicity, our theory applies to more general electron pulse shapes, as the compressed wave function is simply given by the Fourier transform of the incident pulse when considering a perfect lens with an optically imprinted quadratic phase. For illumination confined to a finite spectral range, we must again ensure that the quadratic phase behavior extends over the full duration of the incident wave packet.

Reassuringly, our scheme is robust against variations in the arrival time of the electron, such as those caused by jitter in the electron source. However, temporal coherence is needed to manipulate the incident electron wave function with an optically imprinted phase. This can be achieved by monochromatizing the incident electrons to enhance their longitudinal degree of coherence, though at the cost of reducing the electron beam current. The generation of intense, temporally coherent electron sources remains a challenge that could complement the proposed approach to zeptosecond electron compression.

\section*{Methods}

\subsection{Free-electron interaction with broadband electromagnetic fields}

We consider an electron moving along $z$ with a wave function $\psi(z,t)=\ee^{\ii(q_0z-\mathcal{E}_0t/\hbar)}\phi(z,t)$ characterized by a central wave vector $q_0$, a kinetic energy $\mathcal{E}_0$, and a slowly evolving envelope function $\phi(z,t)$ that is translationally invariant along $(x,y)$ under the assumption of a sufficiently wide electron beam. For small wave vectors $q$ relative to $q_0$, the Taylor expansion of the electron energy $\mathcal{E}_q \approx \mathcal{E}_0 + \hbar v (q-q_0) + \hbar^2(q-q_0)^2/2\me\gamma^3$ with  $\gamma=1/\sqrt{1-v^2/c^2}$, combined with the prescription $q\rightarrow-\ii\hbar\partial_z$, allows us to write the Schr\"odinger equation
\begin{align} \label{Schr} 
\ii\hbar\bigg[\partial_t + v\partial_z - \frac{\ii\hbar}{2\me\gamma^3}\partial_{zz} \bigg] \phi(z,t) = & \Hint(z,t) \phi(z,t),
\end{align}
where $\Hint(z,t)=(ev/c)A_z(z,t)$ introduces the interaction with a classical light field described by a vector potential $\Ab(\rb,t)$. Assuming a small IELS region \cite{paper360} ($\ll\hbar\me v^3\gamma^3/(\Delta\mathcal{E})^2$) that produces relatively small changes in the electron energy ($\Delta \mathcal{E}\ll \mathcal{E}_0$), we can neglect the $\partial_{zz}$ term (nonrecoil approximation) and write the solution of Eq.~(\ref{Schr}) as \cite{paper371}
\begin{align} \label{PINEM1} 
\phi(z,t) =\phi^{\rm inc}(z-vt)\;\ee^{\ii\varphi(z-vt)},
\end{align}
where $\phi^{\rm inc}(z-vt)$ represents the incident electron, while the interaction is encapsulated in the phase function $\varphi(z)=-(1/\hbar v) \int dz'\;\Hint\big[z',(z'-z)/v\big]$. By spectrally decomposing the vector potential as $A_z(z,t)=(1/2\pi)\int d\omega\;A_{z,\omega}(z)\;\ee^{-\ii\omega t}$, the interaction phase can also be written as $\varphi(z)=(\ii/2\pi)\int d\omega\;\beta_\omega\;\ee^{\ii\omega z/v}$ [Eq.~(\ref{phibeta})] in terms of the spectrally resolved coupling coefficient $\beta_\omega=(e/\hbar\omega)\int dz\;E_{z,\omega}(z)\;\ee^{-\ii\omega z/v}$ [Eq.~(\ref{betaw})], where $E_{z,\omega}(z)=(\ii\omega/c)\,A_{z,\omega}(z)$ is the frequency-resolved optical electric field acting along the electron trajectory.

After the interaction, the electron propagates for a long distance $vt$, in which recoil (the $\partial_{zz}$ term) becomes relevant and reshapes the wave function. The initial state $\phi(z,0)$ [Eq.~(\ref{PINEM1}) at $t=0$] evolves as $\phi(z,t)\propto\int dz'\;
\phi^{\rm inc}(z')\;\ee^{\ii\varphi(z')+\ii\pi (\zt-z')^2/vt\lambda_e}$ [Eq.~(\ref{phif}); see Appendix~\ref{appendixA}], where $\lambda_e=2\pi\hbar/(\me v\gamma^3)$, while the optical field enters through $\varphi(z)$ [Eqs.~(\ref{varphibetaw})].

\subsection{Maximal temporal compression}

We consider a normalized incident Gaussian electron wave packet
\begin{align} \label{phii} 
\phi^{\rm inc}(z)\propto\ee^{-z^2/4v^2\sigmat_i^2}
\end{align}
with a temporal standard deviation $\sigmat_i$ in the electron probability density $|\phi^{\rm inc}(z)|^2$. To find the optically imprinted optimum phase profile $\varphi(z)$ that produces a maximum temporal compression at a predetermined free-propagation distance $d$, we calculate the duration of the transmitted wave function in Eq.~(\ref{phif}), $\sigmat_f$. This quantity is defined as $\sigmat_f=(1/v)\sqrt{\mu_2/\mu_0-\mu_1^2/\mu_0^2}$ in terms of the moments of the electron probability density $\mu_n(d)=\int dz\;\ztp^n\,|\phi(z,t)|^2$.

After some algebra, assuming an even function $\varphi(z)$ for simplicity, we find $(\sigmat_f/\sigmat_i)^2=\eta_0^2+\eta_{\rm res}^2$ [Eq.~(\ref{ratio3})], where $\eta_0=\lambda_e d/4\pi v^2\sigmat_i^2$ [Eq.~(\ref{ratio2})] is the minimum possible ratio for the chosen parameters $d$, $v$, and $\sigmat_i$, while
\begin{align} \label{ratio1} 
\eta_{\rm res}^2=(2\eta_0v\sigma_i)^2\int dz \;\big|\phi^{\rm inc}(z)\big|^2 \Big[\varphi'(z)+\frac{2\pi z}{\lambda_ed}\Big]^2
\end{align}
is a residue depending on $\varphi(z)$ (Appendix~\ref{appendixB}).

From Eq.~(\ref{ratio1}), the optimum choice for the phase profile is trivially given by $\varphi(z)=-\pi z^2/\lambda_ed$ [Eq.~(\ref{varphioptimum})] up to an arbitrary constant term. With this choice, integrating Eq.~(\ref{phif}), the wave function reduces to a Gaussian wave packet at all times, and the temporal standard deviation evolves with time as
\begin{align} \label{ratiod} 
\sigmat_t=\sigmat_i\sqrt{\eta_0^2\big(vt/d\big)^2+\big(vt/d-1\big)^2}.
\end{align}
This expression reduces to $\sigma_t=\sigma_i$ before propagation ($t=0$) and $\sigma_t=\eta_0\sigma_i$ at the focusing time $t=d/v$.

\subsection{The uncertainty relation for the post-interaction electron}

Like any Gaussian wave packet, the incident electron wave in Eq.~(\ref{phii}) saturates the uncertainty relation $\sigma_\mathcal{E}\sigma_i\ge\hbar/2$, where $\sigma_\mathcal{E}$ is the energy standard deviation. However, this is not necessarily true after the electron interacts with light. To analyze this effect, we represent the electron energy relative to the central value $\mathcal{E}_0$ by the operator $-\ii\hbar v\partial_z$. After some algebra (see Appendix~\ref{appendixC}), we find
\begin{align} \label{sigmaE2} 
\sigma_\mathcal{E}=\hbar v\sqrt{\frac{1}{4v^2\sigma_i^2}+\int dz\,\big|\phi^{\rm inc}(z)\,\varphi'(z)\big|^2},
\end{align}
which is used together with $\sigma_t$ [Eq.~(\ref{ratiod})] to produce Fig.~\ref{Fig1}c.

\subsection{Practical considerations on the applied optical field}

Inverting Eq.~(\ref{phibeta}) for the phase profile in Eq.~(\ref{varphioptimum}), we find a frequency-resolved coupling coefficient $\beta_\omega=(-\ii\pi\me/\hbar)\,(v^3\gamma^3/d)\,\delta''(\omega)$, which prescribes an unrealistically sharp spectrum emerging from the divergent behavior of $\varphi(z)$ at large $z$. A regularization procedure is thus needed to preserve electron compression with physically viable illumination conditions.

As shown in Fig.~\ref{Fig2}a in the main text, the electron is exposed to external illumination within a limited $z$ region of size $\sim v\sigmat_i$, determined by the incident electron density profile. Therefore, it should be enough that the phase mimics the optimum one only within such a region. To satisfy this condition, we impose the cancellation of the linear term in $z$ within the square brackets of Eq.~(\ref{ratio1}), or equivalently,
\begin{subequations}
\begin{align}
\int \omega d\omega\,\beta_\omega=4\pi^2v/\lambda_ed.  \label{Ibetais1}
\end{align}
In addition, we minimize the required optical pulse energy $\mathcal{E}_{\rm opt}$ to avoid material damage caused by the external illumination. Assuming that the latter is concentrated in a small region compared to the optical wavelengths, Eq.~(\ref{betaw}) implies the proportionality $E_{z,\omega}\propto\omega\beta_\omega$, and consequently,
\begin{align}
\mathcal{E}_{\rm opt}\propto\int d\omega\; |\omega\beta_\omega|^2.  \label{Elight}
\end{align}
\end{subequations}
From various possible schemes explored in Appendices~\ref{appendixD} and \ref{appendixE}, we find that illumination within a finite frequency range $\wmin<\omega<\wmax$ with a moderate ratio $\wmax/\wmin$ is a good practical choice. A constant value of $\beta_\omega=(3\ii\pi\me/\hbar)\,(v^3\gamma^3/\omega_{\rm min}^3d)\,(1-r^3)^{-1}$ with $r=\wmax/\wmin$ maximizes $\mathcal{E}_{\rm opt}$ [Eq.~(\ref{Elight})] and fullfils the condition in Eq.~(\ref{Ibetais1}) (Appendix~\ref{appendixE}). From Eq.~(\ref{ratio1}), the residue becomes $\eta_{\rm res}\approx0.387\,(\wmin\sigmat_i)^2\,(r^5-1)/(r^3-1)$, and thus, $\wmin\sigmat_i\sim\sqrt{\eta_{\rm res}}$ ($\ll1$ for high compression), while the phase profile is
$\varphi(z)=(\ii v\beta_\omega/\pi z)[\sin(\wmax z/v)-\sin(\wmin z/v)]$. For example, for $\sigmat_i\sim100$~fs (attainable using a photoemission source driven by ultrashort laser pulses) and a compression ratio $\eta_{\rm res}<10^{-5}$ (i.e., a minimum final electron pulse duration $\sigma_f<1~$as), the optical spectrum needs to span down to frequencies $\wmin\sim30$~GHz.

\subsection{Electron--light coupling through a metallic slit waveguide}

We focus on an electron passing at a distance $z$ from a metallic slit waveguide of width $a\ll d$, as shown in Fig.~\ref{Fig4}a in the main text. A GHz TEM mode \cite{J99} propagating along the waveguide is reflected at the upper end, exposing the electron to the spilled-out field. Under these conditions, taking $a$ large compared with the skin depth $\sim1\,\mu$m, the metal can be modeled as a perfect conductor, and higher-order waveguide modes can be neglected, leading to a coupling coefficient (Appendix~\ref{appendixF})
\begin{align} \label{betawasymp} 
\beta_\omega\approx\frac{2ea}{\hbar\omega}\;E^{\rm inc}_\omega,
\end{align}
where $E^{\rm inc}_\omega$ is the frequency-resolved incident electric field amplitude in the TEM mode.

\subsection{Optical heating}

From a Poynting-vector analysis (Appendix~\ref{appendixF}), the incident optical energy per pulse is $\mathcal{E}_{\rm opt}=(c\,aL_y/8\pi^2)\int d\omega \,\big|E^{\rm inc}_\omega\big|^2$, where $L_y$ is the waveguide length along the slit direction $y$ (see Fig.~\ref{Fig4}a). Using Eq.~(\ref{betawasymp}) with the optimum constant $\beta_\omega$ within the noted finite frequency range, we find $\mathcal{E}_{\rm opt}/L_y=(3\me^2/16\alpha\hbar)\,(v^6\gamma^6/\omega_{\rm min}^3\,ad^2)\,(r^3-1)^{-1}$. This expression is used to produce Fig.~\ref{Fig4}c.

To estimate an upper bound for the temperature reached by the metal, we calculate the energy density absorbed from the optical pulse at the waveguide walls. This quantity given by $d\mathcal{E}_{\rm abs}/d\rb=(9\me^2/32e^2)\,(v^3\gamma^3/\wmin a d)^2\,(r^4-1)/(r^3-1)^2$, roughly independent of the metal (Appendix~\ref{appendixF}). Assuming an initial room temperature $T_0=300$~K and neglecting heat diffusion, the final wall temperature $T$ is determined from energy conservation as $\gamma T_0^2/2+c_lT_0+d\mathcal{E}_{\rm abs}/d\rb=\gamma T^2/2+c_lT$, where $\gamma$ is the electronic heat coefficient \cite{K1987} and $c_l$ is the lattice heat capacity. For tungsten \cite{S1983_2,L05} ($\gamma\approx106\,$J/m$^3$K$^2$  and $c_l=2.55\times10^6\,$J/m$^3$K$^{-1}$), the resulting $T$ remains well below the melting temperature (3695~K) throughout a broad parameter range in which zeptosecond compression is achieved (see Fig.~\ref{FigS5} in Appendix~\ref{appendixF}).


%

\section*{Acknowledgments}

This work was supported in part by the European Research Council (101141220-QUEFES), the European Commission (101017720-eBEAM and 964591-SMART-electron), the Spanish MCINN (Severo Ochoa CEX2019-000910-S), the Catalan CERCA Program, and Fundaci\'os Cellex and Mir-Puig. X.J. is a visiting student from School of Physics, Peking University, Beijing 100871, China, and acknowledges the Beijing Municipal Natural Science Foundation (grant no. QY23017).

\appendix
\renewcommand{\thefigure}{S\arabic{figure}} 
\setcounter{figure}{0}
\begin{widetext}
\section*{Appendix}

\renewcommand{\theequation}{A\arabic{equation}} 
\renewcommand{\thesection}{A} 
\section{Free-electron interaction with broadband electromagnetic fields}
\label{appendixA}

We consider an electron beam (e-beam) interacting with a classical electromagnetic field described through the vector potential $\Ab(\rb,t)$ in a gauge with vanishing scalar potential. The electron is taken to be highly collimated and characterized by a small momentum and energy spread compared with its central momentum $\hbar\qb_0=\me\vb\gamma$ and kinetic energy $\mathcal{E}_0=\me c^2(\gamma-1)$, where $\vb=v\zz$ is the velocity vector and $\gamma=1/\sqrt{1-v^2/c^2}$. Assuming that the field varies negligibly across the lateral width of the e-beam, we can represent the electron through a time-dependent wave function $\psi(z,t)$ that depends on the longitudinal coordinate $z$ and interacts with the vector potential $\Ab(z,t)\equiv\Ab(0,0,z,t)$ along the e-beam axis $x=y=0$.

Separating a slowly varying envelope $\phi(z,t)$ in the wave function $\psi(z,t)=\ee^{\ii(q_0z-\mathcal{E}_0t/\hbar)}\phi(z,t)$, invoking the second-order Taylor expansion of the electron energy
\begin{align} \nonumber 
\mathcal{E}_q \approx \mathcal{E}_0 + \hbar v (q-q_0) + \hbar^2(q-q_0)^2/2\me\gamma^3
\end{align}
in longitudinal wave vector $q$ relative to $q_0$, and substituting $q$ by $-\ii\partial_z$, we can write the Schr\"odinger equation
\begin{align} \label{Schr} 
\bigg[\ii\hbar\big(\partial_t + v\partial_z \big) + \frac{\hbar^2}{2\me\gamma^3} \partial_{zz} \bigg] \phi(z,t) = & \Hint(z,t) \phi(z,t),
\end{align}
where $\Hint(z,t)=(ev/c)A_z(z,t)$ is the minimal-coupling interaction Hamiltonian, in which $A^2$ terms are neglected, $q$ corrections are dismissed in the prefactor of the vector potential, and the condition $\nabla\cdot\Ab=0$ is considered to be satisfied along the e-beam, provided it does not cross any material interface and is fully contained in vacuum. A derivation of Eq.~(\ref{Schr}) is also possible starting from the Dirac equation \cite{PZ12,paper368}.

When the change in electron energy due to the interaction is small ($\Delta \mathcal{E}\ll \mathcal{E}_0$), the $\partial_{zz}$ term in Eq.~(\ref{Schr}) can be neglected for relatively small propagation distances ($\ll\hbar\me v^3\gamma^3/(\Delta\mathcal{E})^2$ \cite{paper360}). Assuming that $\Hint(z,t)$ is localized in a sufficiently small region such that we can neglect the $\partial_{zz}$ term during the interaction with light (nonrecoil approximation), the transmitted electron wave function admits the analytical solution \cite{paper371}
\begin{align} \label{PINEM1} 
\phi(z,t) =\phi^{\rm inc}(z-vt)\;\ee^{\ii\varphi(z-vt)},
\end{align}
where $\phi^{\rm inc}(z-vt)$ represents the incident electron, while the real phase function
\begin{align} \nonumber 
\varphi(z)&=-\frac{1}{\hbar v} \int dz'\;\Hint\big[z',(z'-z)/v\big]
\end{align}
encapsulates the effects of the interaction. By spectrally decomposing the vector potential as $A_z(z,t)=(1/2\pi)\int d\omega\;A_{z,\omega}(z)\;\ee^{-\ii\omega t}$, the interaction phase can also be written as
\begin{subequations} \label{varphibetaw}
\begin{align} \label{phibeta} 
\varphi(z)=\ii\int\frac{d\omega}{2\pi}\;\beta_\omega\;\ee^{\ii\omega z/v}
\end{align}
in terms of the spectrally resolved coupling coefficient
\begin{align} \label{betaw} 
\beta_\omega=\frac{e}{\hbar\omega}\int dz\;E_{z,\omega}(z)\;\ee^{-\ii\omega z/v},
\end{align}
\end{subequations}
where $E_{z,\omega}(z)=(\ii\omega/c)\,A_{z,\omega}(z)$ is the frequency-space optical electric field.

In the absence of any interaction and neglecting the $\partial_{zz}$ term in Eq.~(\ref{Schr}), the incident wave function must be a solution of $(\partial_t + v\partial_z)\phi^{\rm inc}(z-vt)=0$, and consequently, it depends on $z$ and $t$ only through $z-vt$. The post-interaction wave function $\phi(z,t)$ in Eq.~(\ref{PINEM1}) also depends on $z-vt$ in the nonrecoil approximation. However, recoil becomes relevant in this work when considering large electron propagation distances. Consequently, we need to solve Eq.~(\ref{Schr}) retaining $\partial_{zz}$ and setting $\Hint=0$ in the post-interaction free-propagation region. Starting from an initial state $\phi(z,0)$ given by Eq.~(\ref{PINEM1}) at a time $t=0$ right after the interaction with light, and using the momentum representation
\begin{align} \nonumber 
\phi(z,t)=\int\frac{dq}{2\pi}\;\phi_q(t)\;\ee^{\ii qz},
\end{align}
the evolution of the wave vector components is then determined by the equation $\partial_t\phi_q(t)=-\ii qv(1+q\lambda_e/4\pi)\phi_q(t)$ with
\begin{align} \nonumber 
\lambda_e=\frac{2\pi\hbar}{\me v\gamma^3}
\end{align}
(i.e., the de Broglie wavelength of the electron corrected by a factor $1/\gamma^2$), whose solution reduces to
\begin{align} \nonumber 
\phi(z,t)=\int\frac{dq}{2\pi}\;\ee^{\ii q(z-vt)}\ee^{-\ii q^2\lambda_e vt/4\pi}
\int dz' \;\ee^{-\ii qz'}\;\phi^{\rm inc}(z')\;\ee^{\ii\varphi(z')},
\end{align}
where the rightmost integral stands for the momentum components of the $t=0$ wave function $\phi(z,0)$. The $q$ integral in this equation can be performed analytically, and we finally obtain
\begin{align} \label{phif} 
\phi(z,t)=\frac{\ee^{-\ii\pi/4}}{\sqrt{vt\lambda_e}}\int dz'\;
\phi^{\rm inc}(z')\;\ee^{\ii\varphi(z')+\ii\pi(\zt-z')^2/vt\lambda_e}.
\end{align}
The interaction with light thus transforms an incident electron wave function $\phi^{\rm inc}(z-vt)$ into the one given in Eq.~(\ref{phif}), where the optical field enters through the phase in Eqs.~(\ref{varphibetaw}).

\renewcommand{\theequation}{B\arabic{equation}} 
\renewcommand{\thesection}{B} 
\section{Maximal temporal compression}
\label{appendixB}

In what follows, we consider a normalized incident Gaussian electron wave packet 
\begin{align} \label{phii} 
\phi^{\rm inc}(z)=\frac{\ee^{-z^2/4v^2\sigmat_i^2}}{(2\pi v^2\sigmat_i^2)^{1/4}}
\end{align}
with a temporal standard deviation $\sigmat_i$ in the electron probability density $|\phi^{\rm inc}(z)|^2$. We intend to find the optimum optically imprinted phase profile $\varphi(z)$ that produces a maximum temporal compression of the wave function according to Eq.~(\ref{phif}). The specific form of $\varphi(z)$ might depend on the incident wave function, but the general procedure to obtain it remains the same.

We calculate the temporal standard deviation of the transmitted wave function given in Eq.~(\ref{phif}) to assess the degree of electron compression. This quantity is given by $\sigmat_f=(1/v)\sqrt{\mu_2/\mu_0-\mu_1^2/\mu_0^2}$ in terms of the moments of the electron probability density. After some algebra, the $n$-th moment is found to be
\begin{align} \nonumber 
\mu_n(d)=\int dz\;\ztp^n\,|\phi(z,t)|^2=(\ii\,\lambda_ed/2\pi)^n
\int dz \;\phi^{\rm inc}(z) \;\ee^{\ii\varphi(z)+\ii\pi z^2/\lambda_ed}
\frac{\partial^n}{\partial z^n}\Big\{\big[\phi^{\rm inc}(z)\big]^* \;\ee^{-\ii\varphi(z)-\ii\pi z^2/\lambda_ed}\Big\}
\end{align}
as a function of free-propagation distance $d=vt$. For the specific incident wave function in Eq.~(\ref{phii}), the first three moments reduce to
\begin{align} \nonumber 
\begin{aligned}
\mu_0(d)&=\int dz \;\big|\phi^{\rm inc}(z)\big|^2=1, \\
\mu_1(d)&=\frac{\lambda_ed}{2\pi}\int dz \;\big|\phi^{\rm inc}(z)\big|^2 \;\varphi'(z), \\
\mu_2(d)&=\Big(\frac{\lambda_ed}{2\pi}\Big)^2\bigg\{\frac{1}{(2v\sigmat_i)^2}+\int dz \;\big|\phi^{\rm inc}(z)\big|^2 \Big[\varphi'(z)+\frac{2\pi z}{\lambda_ed}\Big]^2\bigg\}.
\end{aligned}
\end{align}
To find a phase function $\varphi(z)$ that minimizes $\sigmat_f$, we apply the Euler--Lagrange equation and derive the condition $\varphi'(z)=-2\pi z/\lambda_ed+Q$, where $Q$ is an irrelevant $z$-independent term, as it cancels when this expression is inserted in the definition of $\sigmat_f$, yielding $\sigma_f=\lambda_ed/4\pi v^2\sigma_i$. We can thus write an optimum phase
\begin{align} \label{varphiz} 
\varphi(z)=\varphi_{\rm optimum}(z)\equiv-\frac{\pi z^2}{\lambda_ed}
\end{align}
up to an arbitrary constant phase for compression at a propagation distance $d$. The $z^2$ dependence of the phase is a reminiscence of the quadratic profile introduced by optical lenses in the paraxial approximation, therefore indicating that we are dealing with a form of lensing, but in the temporal domain, with short free-propagation distances $d$ and large electron velocities favoring compression. We elaborate further on this concept in Fig.~\ref{Fig3} of the main text.

As discussed below, deviations from the profile in Eq.~(\ref{varphiz}) are introduced when considering attainable illumination conditions. For simplicity, we assume even functions $\varphi(z)$, for which we have $\mu_1=0$, while the ratio of standard deviations becomes
\begin{align} \label{ratio1} 
\frac{\sigmat_f}{\sigmat_i}=\eta_0
\;\sqrt{1+(2v\sigmat_i)^2\int dz \;\big|\phi^{\rm inc}(z)\big|^2 \Big[\varphi'(z)+\frac{2\pi z}{\lambda_ed}\Big]^2},
\end{align}
where
\begin{align} \label{ratio2} 
\eta_0\equiv\frac{\lambda_ed}{4\pi v^2\sigmat_i^2}
= \frac{\hbar d}{2\me v^3\gamma^3\sigmat_i^2}
\end{align}
is the minimum value obtained for the phase in Eq.~(\ref{varphiz}), which is, therefore, the minimum compression ratio that can be obtained for a given combination of propagation distance $d$, electron velocity $v$, and incident pulse duration $\sigmat_i$.

With the phase optimized in Eq.~(\ref{varphiz}) for a propagation distance $d$, the temporal evolution of electron probability in Eq.~(\ref{phif}) becomes a Gaussian $|\phi(z,t)|^2=(\sqrt{2\pi} v\sigmat_t)^{-1}\ee^{-\ztp^2/2v^2\sigmat_t^2}$ with a time-dependent temporal standard deviation
\begin{align} \label{ratiod} 
\sigmat_t=\sigmat_i\sqrt{\eta_0^2\Big(\frac{vt}{d}\Big)^2+\Big(\frac{vt}{d}-1\Big)^2}.
\end{align}
This expression reduces to $\sigma_t=\sigma_i$ before propagation ($t=0$) and to $\sigma_t=\eta_0\sigma_i$ at the focusing time $t=d/v$. Interestingly, Eq.~(\ref{ratiod}) predicts a maximum compression $\eta_0/\sqrt{1+\eta_0^2}$ (i.e., a minimum of $\sigma_t/\sigma_i$) at a distance $vt=d/(1+\eta_0^2)$ slightly shorter than $d$.

\renewcommand{\theequation}{C\arabic{equation}} 
\renewcommand{\thesection}{C} 
\section{The uncertainty relation applied to the post-interaction electron}
\label{appendixC}

Like any Gaussian wave packet, the incident electron wave in Eq.~(\ref{phii}) saturates the uncertainty relation $\sigma_\mathcal{E}\sigma_i\ge\hbar/2$, where $\sigma_\mathcal{E}$ is the energy standard deviation (i.e., the equal sign applies). However, this is no longer true after the electron interacts with light. To analyze this effect, we represent the electron energy relative to the central value $\mathcal{E}_0$ by the operator $-\ii\hbar v\partial_z$, assuming that the spatial extension of the electron wave packet is short enough to adopt the nonrecoil approximation. Therefore, we can calculate $\sigma_\mathcal{E}$ as
\begin{align} \label{sigmaE1} 
\sigma_\mathcal{E}=\hbar v\sqrt{\frac{\int dz\,|\partial_z\phi(z,t)|^2}{\int dz\,|\phi(z,t)|^2}+\bigg(\frac{\int dz\,\phi^*(z,t)\partial_z\phi(z,t)}{\int dz\,|\phi(z,t)|^2}\bigg)^2},
\end{align}
which readily yields $\sigma_\mathcal{E}=\hbar/2\sigma_i$ when substituting $\phi^{\rm inc}(z-vt)$ [Eq.~(\ref{phii})] for $\phi(z,t)$. Considering the post-interaction wave function in Eq.~(\ref{PINEM1}), we have $\partial_z\phi(z,t)=\phi^{\rm inc}(\zt)\,[\ii\varphi'(\zt)-\zt/2(v\sigma_i)^2]$ and, using the fact that $\phi^{\rm inc}(\zt)$ is normalized and $\varphi(z)$ is assumed to be an even function, Eq.~(\ref{sigmaE1}) reduces to
\begin{align} \label{sigmaE2} 
\sigma_\mathcal{E}=\hbar v\sqrt{\frac{1}{4v^2\sigma_i^2}+\int dz\,\big|\phi^{\rm inc}(z)\,\varphi'(z)\big|^2}.
\end{align}
This quantity remains constant after the interaction because the electron spectrum does not change during free-space propagation. In the main text, we use Eq.~(\ref{sigmaE2}) in combination with $\sigma_f$ in Eq.~(\ref{ratio3}) to calculate $\sigma_\mathcal{E}\sigma_f$ and compare the result with the lower bound $\hbar/2$ imposed by the uncertainty principle.

\renewcommand{\theequation}{D\arabic{equation}} 
\renewcommand{\thesection}{D} 
\section{Practical considerations on the spectral profile of the optical field}
\label{appendixD}

Inverting Eq.~(\ref{phibeta}), we can determine the spectral decomposition of the coupling coefficient required to obtain a desired phase profile $\varphi(z)$ as
\begin{align} \label{betawphiz} 
\beta_\omega=-\frac{\ii}{v}\int dz\;\varphi(z)\;\ee^{-\ii\omega z/v}.
\end{align}
When plugging Eq.~(\ref{varphiz}) into this expression, we obtain $\beta_\omega=(-2\pi^2\ii v^2/\lambda_ed)\,\delta''(\omega)$ (i.e., proportional to the second derivative of the $\delta$-function), which prescribes an unrealistically sharp spectrum emerging from the divergent behavior of $\varphi(z)$ at large $z$. We thus need to find a regularization procedure that preserves a sufficiently large degree of electron compression while rendering the external illumination spectrum attainable. Fortunately, as shown in Fig.~\ref{Fig2}a in the main text, the electron is exposed to external illumination within a limited $z$ region of size $\sim v\sigmat_i$ determined by the incident electron density profile. Therefore, the phase must mimic the optimum one only within that region. We intend to exploit this idea and find feasible optimum illumination conditions.

The maximum light energy needed to produce the desired optical spectral shape should be maintained within reasonable limits. In addition, light modulation can be challenging if the spectral range is too large, so we consider illumination comprising frequency components within a reasonably narrow range. In what follows, we explore three different regularization schemes that are evaluated according to their ability to meet these criteria (i.e., small frequency range and minimum possible optical energy for a given degree of electron temporal compression):
\begin{itemize}
\item[({\it i})] Illumination within a finite frequency range $\wmin<\omega<\wmax$ for a not-too-large ratio $\wmax/\wmin$ (Sec.~\ref{scheme1}).
\item[({\it ii})] Real-space phase profile $\varphi(z)$ corrected by a Gaussian decay at large $|z|$ (Sec.~\ref{scheme2}).
\item[({\it iii})] Gaussian spectral profile adjusted to produce the best possible compression (Sec.~\ref{scheme3}).
\end{itemize}
Scheme ({\it i}) is appealing because it relies on a finite frequency range and requires a nearly optimum light energy (see Sec.~\ref{schemecomparison}), so we use it to obtain the results presented in the main text.

The optical energy $\mathcal{E}_{\rm opt}$ associated with the optimum illumination conditions depends on the electron--light coupling configuration. Still, we can estimate its dependence on illumination and electron parameters from the relation between the electric field and the coupling coefficient in Eq.~(\ref{betaw}). For example, with an extended interaction region (fixed length $L$) in which all frequency components are made to be in phase (i.e., $E_{z,\omega}\propto\ee^{\ii\omega z/v}$), we have $E_{z,\omega}\propto\omega\beta_\omega/L$. However, for a localized coupling region (e.g., a thin film or a tip), low $\omega$'s act over a larger spatial extension $L\propto1/\omega$, increasing with the light wavelength, so we find a scaling closer to $E_{z,\omega}\sim\omega^2\beta_\omega$. To discuss these possibilities qualitatively, we consider the approximate relation
\begin{align} \label{Fn1} 
\mathcal{E}_{\rm opt}\approx C\int d\omega\; |\omega^n\beta_\omega|^2,
\end{align}
with $n=1,2$. Here, $C$ is a frequency-independent constant determined by the specific geometry under consideration. Incidentally, for geometries that require illumination of a finite spatial region (e.g., in Sec.~\ref{TEMWG}), $C$ has units of (energy)$\times$(time)$^{2n-1}$. However, for configurations involving the illumination of extended areas (see, for example, Secs.~\ref{PECfilm}), it makes sense to talk about the incident optical energy per unit of surface area (i.e., the fluence), as $C$ is proportional to the illuminated area. For simplicity, we refer to the optical energy in all cases.

In brief, we intend to explore different spectral light profiles that can produce the desired degree of electron temporal compression as quantified by the ratio of incident-to-final electron pulse durations [Eq.~(\ref{ratio1})] while limiting the optical energy [Eq.~(\ref{Fn1})] within affordable limits. We find it convenient to rewrite Eqs.~(\ref{ratio1}) and (\ref{Fn1}) in dimensionless units by changing the variables of integration to
\begin{align}
&\theta=z/v\sigmat_i, \nonumber\\
&\wt=\omega\sigmat_i, \nonumber
\end{align}
normalized through the temporal width of the incident electron $\sigmat_i$. Using the explicit expression for the incident wave function in Eq.~(\ref{phii}) and writing $\varphi(z)$ in terms of $\beta_\omega$ through Eq.~(\ref{phibeta}), we obtain, after some straightforward algebra,
\begin{subequations} 
\begin{align}  
&\frac{\sigmat_f}{\sigmat_i}=\sqrt{\eta_0^2+\eta_{\rm res}^2}, \label{ratio3}\\
&\mathcal{E}_{\rm opt}\approx\frac{\pi^2C}{\eta_0^2\sigmat_i^{2n-1}}\;F_n, \label{Fn2}
\end{align}
where $\eta_0$ [Eq.~(\ref{ratio2})] determines the maximum compression for a given choice of $v$, $\sigmat_i$, and $d$; we define
\begin{align} \label{ratio4} 
\eta_{\rm res}=\frac{1}{(2\pi)^{1/4}}\sqrt{\int d\theta \;\ee^{-\theta^2/2} \bigg[\theta-\int d\wt \;\wt\,f_{\wt}\,\sin(\theta\wt)\bigg]^2}
\end{align}
as a residual contribution to the compression ratio associated with imperfect optimization in the chosen regularization scheme; the optical spectrum enters the optical energy $\mathcal{E}_{\rm opt}$ through the dimensionless parameter
\begin{align} \label{Fn3} 
F_n=\int d\wt\; \wt^{2n} f_{\wt}^2\,;
\end{align}
and the optical coupling coefficient is rewritten as
\begin{align} \label{betawfw} 
\beta_\omega=-\frac{\ii\pi\sigmat_i}{\eta_0} f_{\wt}
\end{align}
\end{subequations}
in terms of a dimensionless function $f_{\wt}$. Since $\varphi(z)$ is a real function of $z$ that we assumed to be even, Eq.~(\ref{betawphiz}) implies that $f_{\wt}$ is also real and even in $\wt=\omega\sigmat_i$. With the aforementioned change of variables, the optimum phase reduces to
\begin{align} \label{varphiopt} 
\varphi_{\rm optimum}(z)=-\frac{\theta^2}{4\eta_0},
\end{align}
which is inversely proportional to the maximum compression fraction $\eta_0$.

\renewcommand{\theequation}{E\arabic{equation}} 
\renewcommand{\thesection}{E} 
\section{Phase regularization schemes}
\label{appendixE}

\subsection{Scheme ({\it i}): Optimum illumination within a finite frequency range}
\label{scheme1}

We consider illumination with a frequency decomposition in a finite range defined by $\wmin<\omega<\wmax$. Optimum compression is obtained if the spectral function $f_{\wt}$ is shaped such that the $\wt$ integral in Eq.~(\ref{ratio4}) equals $\theta$. As a practical strategy, rather than attempting a rigorous optimization of $f_{\wt}$ that minimizes the optical energy for a given fixed value of the residual fraction $\eta_{\rm res}$, we examine the Taylor expansion of the $\wt$ integral around $\theta=0$,
\begin{align} \label{Taylor} 
\int d\wt \;\wt\,f_{\wt}\,\sin(\theta\wt)=\theta \int d\wt \;\wt^2\,f_{\wt}-\frac{\theta^3}{6} \int d\wt \;\wt^4\,f_{\wt}+\cdots,
\end{align}
and impose the condition that the linear term cancels $\theta$ inside the squared brackets of Eq.~(\ref{ratio4}). We thus look for a function $f_{\wt}$ that minimizes the optical energy [i.e., $F_n$, see Eq.~(\ref{Fn3})], subject to the condition $\int d\wt \;\wt^2\,f_{\wt}=1$. Applying the Euler--Lagrange equation, we find
\begin{align} \label{fwtscheme1} 
f_{\wt}=\left\{
\begin{matrix}
\lambda_n\,\wt^{2-2n} & (\wmin<\omega<\wmax), \\
\\
0 & \quad\quad\quad ({\rm otherwise}),
\end{matrix}
\right.
\end{align}
where
\begin{align} \nonumber 
\lambda_n=\frac{5/2-n}{\wt_{\rm max}^{5-2n}-\wt_{\rm min}^{5-2n}}
\end{align}
is a Lagrange multiplier determined from the noted condition. The real-space phase profile is then obtained by inserting Eq.~(\ref{fwtscheme1}) into Eqs.~(\ref{phibeta}) and (\ref{betawfw}), leading to $\varphi(z)=(1/2\eta_0)\int d\wt \,f_{\wt}\,\ee^{\ii\theta\wt}$, which reduces to $\varphi(z)=(\lambda_1/\eta_0\theta)[\sin(\theta\wt_{\rm max})-\sin(\theta\wt_{\rm min})]$ for $n=1$ and $\varphi(z)=(\lambda_2/\eta_0)\int_{\wt_{\rm min}}^{\wt_{\rm max}} d\wt \;\wt^{-2}\cos(\theta\wt)$ for $n=2$. In addition, the optical-energy parameter reduces to $F_n=\lambda_n$, while the residual compression fraction in Eq.~(\ref{ratio4}) is obtained by solving the $\wt$ integral analytically,\footnote{In the evaluation of $\eta_{\rm res}$, the integral $\int d\wt \;\wt\,f_{\wt}\,\sin(\theta\wt)=2[I(\theta,\wt_{\rm max})-I(\theta,\wt_{\rm min})]$ in Eq.~(\ref{ratio4}) can be calculated analytically for $f_{\wt}$ given by Eq.~(\ref{fwtscheme1}). We find $I(\theta,\wt)=(\lambda_1/\theta^2)[\sin(\theta\wt)-\theta\wt\,\cos(\theta\wt)]$ for $n=1$ and $I(\theta,\wt)=\lambda_2{\rm Si}(\theta\wt)$ for $n=2$, where ${\rm Si}(x)$ is the sine integral function.} followed by numerical integration over $\theta$.

In this regularization scheme, $\eta_{\rm res}$ and $F_n$ are functions of the dimensionless parameters $\wmin\sigmat_i$ and $r=\wmax/\wmin$. We can obtain an approximation to $\eta_{\rm res}$ by retaining the $\theta^3$ term in the $\wt$ integral [Eq.~(\ref{Taylor})], as the $\theta$ term cancels by construction. Then, Eq.~(\ref{ratio4}) leads to the analytical result
\begin{subequations} \label{scheme1summary} 
\begin{align} \label{scheme1approxetares} 
\eta_{\rm res}&\approx\sqrt{\frac{5}{12}}\bigg(\frac{5/2-n}{7/2-n}\bigg)\bigg(\frac{r^{7-2n}-1}{r^{5-2n}-1}\bigg)\big(\wmin\sigmat_i\big)^2 \nonumber\\
&\approx\big(\wmin\sigmat_i\big)^2\times\left\{
\begin{matrix}
0.387\,(r^5-1)/(r^3-1) & \quad\quad\quad (n=1), \\
\\
0.215\,(r^2+r+1) & \quad\quad\quad (n=2).
\end{matrix}
\right.
\end{align}
This approximation works extremely well within a large range of $r$ values for small $\eta_{\rm res}$ (see Fig.~\ref{FigS1}a). Using the same notation, we have
\begin{align} \label{scheme1Fn} 
F_n=\frac{5/2-n}{r^{5-2n}-1}\,\big(\wmin\sigmat_i\big)^{2n-5}=\left\{
\begin{matrix}
\big[1.5/(r^3-1)\big]\,\big(\omega_{\rm min}\sigmat_i\big)^{-3} & \quad\quad\quad (n=1), \\
\\
\big[0.5/(r-1)\big]\,\big(\omega_{\rm min}\sigmat_i\big)^{-1} & \quad\quad\quad (n=2)
\end{matrix}
\right.
\end{align}
\end{subequations}
for the optical-energy parameter.

From Eq.~(\ref{scheme1approxetares}), we find the scaling $\wmin\sigmat_i\sim\sqrt{\eta_{\rm res}}$ (i.e., $\wmin\sigmat_i\ll1$ to achieve a high compression). The optical spectral range is thus determined by the temporal extension of the incident electron wave packet, and for example, for $\sigmat_i\sim100$~fs (an initial electron pulse duration attainable using a photoemission source driven by ultrashort laser pulses) and a compression ratio $\eta_{\rm res}<10^{-5}$ (i.e., a final electron pulse duration $\sigma_f<1~$as), the optical spectrum needs to span down to frequencies $\wmin\sim30$~GHz.

Combining Eqs.~(\ref{scheme1summary}), we can eliminate $\wmin\sigmat_i$ and obtain the optical-energy parameter $F_n$ as a function of $\eta_{\rm res}$ and the spectral ratio $r$, as plotted in Fig.~\ref{FigS2}, where we compare different regularization schemes (see Sec.~\ref{schemecomparison}).

\begin{figure*}[htb]
\centering \includegraphics[width=0.9\textwidth]{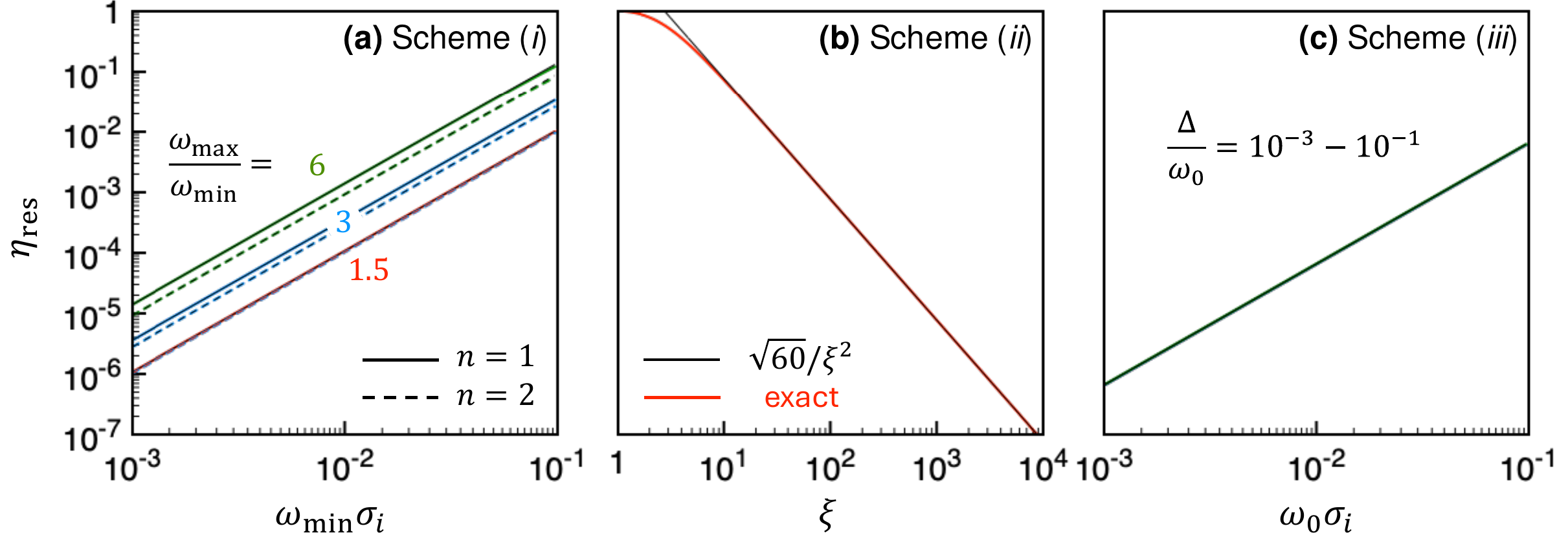}
\caption{{\bf Comparison of different phase regularization schemes: Analytical approximation vs numerically calculated residual compression ratio.} We plot the residual compression fraction $\eta_{\rm res}$ given by the analytical approximations in Eqs.~(\ref{scheme1approxetares}), (\ref{scheme2approxetares}), and (\ref{scheme3approxetares}) (thin black curves) compared with the results of rigorous numerical integration (thick curves). The approximated curves are almost indistinguishable from the numerical integration. Panels (a)--(c) correspond to the schemes discussed in Secs.~\ref{scheme1}--\ref{scheme3}, respectively. In (a), we show results for $n=1$ (solid curves) and $n=2$ (broken curves) [see Eq.~(\ref{Fn1}) for a definition of $n$]. Curves in panels (b) and (c) are independent of $n$.}
\label{FigS1}
\end{figure*}

\subsection{Scheme ({\it ii}): Gaussian real-space correction}
\label{scheme2}

An alternative strategy to eliminate the divergent behavior of $\varphi(z)$ at large $|z|$ consists in superimposing a Gaussian profile on the optimum phase given by Eq.~(\ref{varphiopt}), that is,
\begin{align} \label{correctedphase1} 
\varphi(z)=\varphi_{\rm optimum}(z)\,\ee^{-z^2/v^2\sigmat_l^2}=-\frac{\theta^2}{4\eta_0}\,\ee^{-\theta^2/\xi^2},
\end{align}
where $\sigmat_l$ stands for the Gaussian duration of the optical field in the electron--light interaction region, and we define the ratio $\xi=\sigmat_l/\sigmat_i$. Inserting Eq.~(\ref{correctedphase1}) into Eq.~(\ref{betawphiz}) and adopting the definition in Eq.~(\ref{betawfw}), we find
\begin{align} \label{fwt2} 
f_{\wt}=(\xi^3/8\sqrt{\pi})\,\big(\wt^2\xi^2/2-1\big)\,\ee^{-\wt^2\xi^2/4}.
\end{align}
We then plug this expression into Eq.~(\ref{ratio4}) and find the analytical result
\begin{align}
\eta_{\rm res}=\sqrt{1+\frac{\xi^3}{(\xi^2+4)^{\frac{3}{2}}}\bigg[1-\frac{6}{\xi^2+4}+\frac{15}
{(\xi^2+4)^2}\bigg]-\frac{2\xi^3}{(\xi^2+2)^{\frac{3}{2}}}\bigg[1-\frac{3}{\xi^2+2}\bigg]}
\end{align}
for the residual compression ratio, which enters the full compression ratio through Eq.~(\ref{ratio3}). Reassuringly, the residual phase vanishes as
\begin{subequations} \label{scheme2summary} 
\begin{align} \label{scheme2approxetares} 
\eta_{\rm res}\approx\frac{\sqrt{60}}{\xi^2}
\end{align}
in the $\xi\gg1$ limit (see Fig.~\ref{FigS1}b), so we recover the perfect compression result $\sigmat_f/\sigmat_i\rightarrow\eta_0$. However, this limit can demand large values of $\xi$: an extended Gaussian is needed to maintain the optimum phase in the interaction region. For example, for a maximum compression fraction $\eta_0\sim10^{-5}$, the condition $\eta_{\rm res}=\eta_0$ is satisfied for $\xi\sim880$.

With the spectral profile in Eq.~(\ref{fwt2}), the optical-energy parameter [Eq.~(\ref{Fn3})] becomes $F_n=(\xi^{5-2n}/\pi2^{6-n-1/2})(n^2+3/4)\Gamma(n+1/2)$ in terms of the gamma function, and in particular,
\begin{align} \label{scheme2Fn} 
F_n=\frac{1}{128\sqrt{2\pi}}\times\left\{
\begin{matrix}
7\,\xi^3 & \quad\quad\quad (n=1), \\
\\
57\,\xi & \quad\quad\quad (n=2)
\end{matrix}
\right.
\end{align}
\end{subequations}
for $n=1,2$. In this regularization scheme, $\eta_{\rm res}$ and $F_n$ are functions of a single parameter $\xi=\sigmat_l/\sigmat_i$. Upon substitution in Eq.~(\ref{scheme2Fn}), we obtain the optical-energy parameter that is required as a function of the targeted residual compression ratio (thin horizontal lines in Fig.~\ref{FigS2}).

\subsection{Scheme ({\it iii}): Gaussian spectral optical profile}
\label{scheme3}

We explore yet another practical regularization scheme to avert the pathological behavior of $\beta_\omega$: a Gaussian spectral profile that mimics the optimum phase in Eq.~(\ref{varphiz}) over a sufficiently large spatial range. More precisely, we consider the real, even spectral function
\begin{align} \label{fwtscheme3} 
f_{\wt}=s\Big[\ee^{-(\wt-\wt_0)^2/\Dt^2}+\ee^{-(\wt+\wt_0)^2/\Dt^2}\Big],
\end{align}
which, using Eqs.~(\ref{phibeta}) and (\ref{betawfw}), renders the phase profile $\varphi(z)=(s\sqrt{\pi}\Dt/\eta_0)\,\cos(\theta\wt_0)\,\ee^{-\theta^2\Dt^2/4}$. The spectrum is characterized by a scaled central frequency $\wt_0=\omega_0\sigmat_i$ and width $\Dt=\Delta\sigmat_i$. Like in Sec.~\ref{scheme1}, the overall optical amplitude [i.e., the coefficient $s$ in Eq.~(\ref{fwtscheme3})] is determined by the condition $\int d\wt \;\wt^2\,f_{\wt}=1$, which guarantees that the linear $\theta$ term vanishes inside the square brackets of Eq.~(\ref{ratio4}). We find
\begin{align} \label{s_coef} 
s=\frac{1}{\sqrt{\pi}\Dt(\Dt^2+2\wt_0^2)}.
\end{align}
In this regularization scheme, $\eta_{\rm res}$ and $F_n$ are functions of the parameters $\omega_0\sigmat_i$ and $f=\Delta/\omega_0$. The compression fraction $\eta_{\rm res}$ [Eq.~(\ref{ratio4})] is obtained by solving the $\wt$ integral analytically,\footnote{Taking $f_{\wt}$ from Eq.~(\ref{fwtscheme1}), we obtain $\int d\wt \;\wt\,f_{\wt}\,\sin(\theta\wt)=s\sqrt{\pi}\Dt\,\big[2\wt_0\sin(\theta\wt_0)+\theta\Dt^2\cos(\theta\wt_0)\,]\,\ee^{-\theta^2\Dt^2/4}$.} followed by numerical integration over $\theta$. In analogy to Sec.~(\ref{scheme1}), an approximation to $\eta_{\rm res}$ can also be obtained by retaining the $\theta^3$ term in the $\wt$ integral [Eq.~(\ref{Taylor})]:
\begin{subequations} \label{scheme3summary} 
\begin{align} \label{scheme3approxetares} 
\eta_{\rm res}&\approx\sqrt{\frac{5}{48}}\bigg(\frac{4+12f^2+3f^4}{2+f^2}\bigg)\big(\omega_0\sigmat_i\big)^2,
\end{align}
which compares very well with the full numerical result (see Fig.~\ref{FigS1}c). In addition, the optical-energy parameter $F_n$ is calculated by inserting Eq.~(\ref{fwtscheme3}) into Eq.~(\ref{Fn3}), leading to the analytical expressions
\begin{align} \label{scheme3Fn} 
F_n=\sqrt{\frac{2}{\pi}}\frac{1}{f(2+f^2)^2}\times\left\{
\begin{matrix}
\big[1+(f^2/4)\big(1+\ee^{-2/f^2}\big)\big]\,\big(\omega_0\sigmat_i\big)^{-3} & \quad\quad\quad (n=1), \\
\\
\big[1+3f^2/2+(3f^4/16)\big(1+\ee^{-2/f^2}\big)\big]\,\big(\omega_0\sigmat_i\big)^{-1} & \quad\quad\quad (n=2)
\end{matrix}
\right.
\end{align}
\end{subequations}
for $n=1,2$.

A Gaussian spectral profile effectively limits the optical spectrum to a relatively narrow range, so it is not surprising that we find an analogous scaling $\omega_0\sigmat_i\sim\sqrt{\eta_{\rm res}}$ as in scheme ({\it i}) [Sec.~(\ref{scheme1})]. Again, combining Eqs.~(\ref{scheme3summary}), we can eliminate $\omega_0\sigmat_i$ and obtain the optical-energy parameter $F_n$ as a function of $\eta_{\rm res}$ and $f$ (see Fig.~\ref{FigS2}b and Sec.~\ref{schemecomparison}).

\begin{figure*}[htb]
\centering \includegraphics[width=0.6\textwidth]{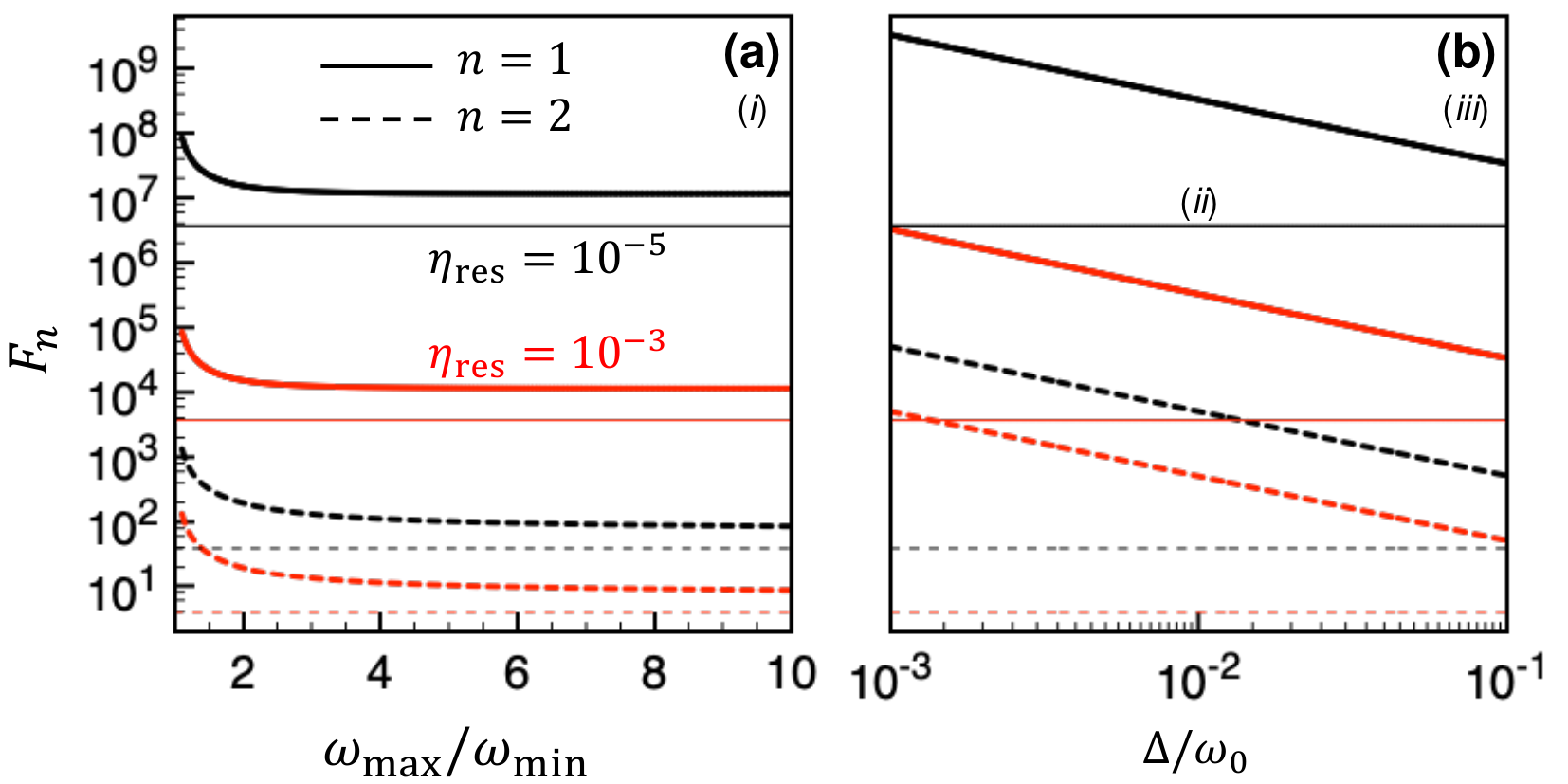}
\caption{{\bf Comparison of different phase regularization schemes: Optical-energy parameter for a fixed residual compression ratio.} We plot the optical-energy parameters $F_1$ (solid curves) and $F_2$ (broken curves) for fixed residual compression ratios $\eta_{\rm res}=10^{-3}$ (red curves) and $10^{-5}$ (black curves) using (a) scheme ({\rm i}) and (b) scheme ({\rm iii}). The results obtained from the scheme ({\rm ii}) are shown in both panels for comparison (thin horizontal lines).}
\label{FigS2}
\end{figure*}

\subsection{Critical comparison of different regularization schemes}
\label{schemecomparison}

In Fig.~\ref{FigS2}, we plot the optical-energy parameters $F_1$ and $F_2$ [see Eqs.~(\ref{Fn1}) and (\ref{Fn2})] obtained from the above regularization schemes for two fixed values of the residual compression ratio $\eta_{\rm res}=10^{-3}$ and $10^{-5}$. Scheme ({\it ii}) systematically requires the smallest optical energies (thin horizontal lines). Still, it involves a wide frequency range covering down to $\omega=0$ [see Eq.~(\ref{fwt2})], which can be challenging to implement. However, the compression ratio dramatically increases when we introduce a frequency cutoff to sort this problem out. Scheme ({\it iii}) demands a similar level of optical energy for a Gaussian spectral shape of relative width $\Delta/\omega_0\sim1$ (Fig.~\ref{FigS2}b), but then, the spectrum takes a substantial amplitude down to $\omega=0$. Unfortunately, in this scheme, a sufficiently narrower Gaussian profile (as required to avert $\omega=0$) increases the optical energy well above the other two schemes. Finally, scheme ({\it i}) intrinsically imposes a finite frequency range and, for a spectral ratio $\wmax/\wmin>4$, the required optical energy is just a factor of $2-3$ larger than for scheme ({\it ii}) (Fig.~\ref{FigS2}a). Therefore, we postulate scheme ({\it i}) as a practical approach that features a finite frequency range and a nearly optimum optical-energy parameter.

\begin{figure*}[htb]
\centering \includegraphics[width=0.4\textwidth]{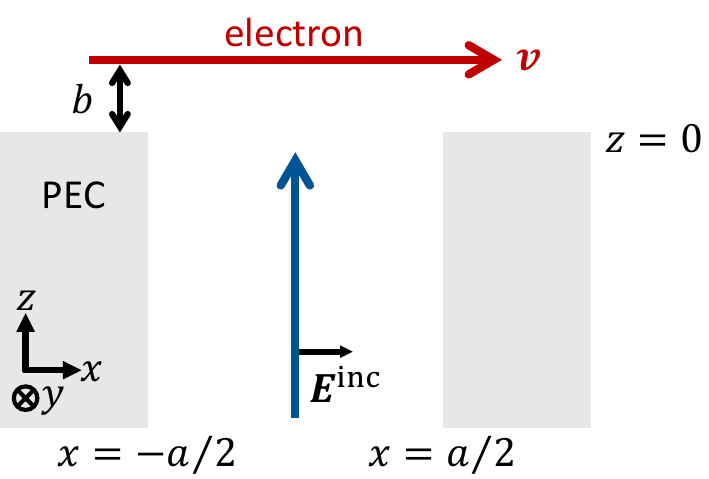}
\caption{{\bf Electron coupling to GHz fields through a TEM waveguide.} A TEM wave is propagating upward in a slit of width $a$ perforated in a perfect electric conductor (PEC). There is reflection at the waveguide end ($z=0$ plane) and spill out of the field on the $z>0$ region, where an electron is passing parallel to the upper PEC surface.}
\label{FigS3}
\end{figure*}

\renewcommand{\theequation}{F\arabic{equation}} 
\renewcommand{\thesection}{F} 
\section{Electron passing near the end of a metallic TEM waveguide}
\label{TEMWG}
\label{appendixF}

In the main text, we consider GHz optical fields, whose wavelengths ($\sim$ a few cm) are much larger than the free electron propagation distance $d\sim\,$mm required to achieve large electron temporal compression. To make the interaction region much smaller than $d$, we consider a metallic waveguide supporting TEM modes, as shown in Fig.~\ref{FigS3}. In the sketch, a GHz TEM wave propagates upward and is partially reflected at the aperture. The field spills out into the outer vacuum region traversed by the electron. For simplicity, in the following analysis, we consider a slit waveguide of width $a$, translational symmetry along the direction $y$ perpendicular to the sketch, and infinite extension toward $z<0$. The upper metal surface is at $z=0$ and the waveguide is defined by the side walls at $x=\pm a/2$. The metal is modeled as a perfect electric conductor (PEC), which should be a reasonable assumption for low-frequency fields and large $a$ compared to the skin depth (see below). In this configuration, the optical electric field has a vanishing component along $y$. We first derive the electric field right outside the waveguide and then calculate the electron--light coupling coefficient through Eq.~(\ref{betaw}).

\subsection{Optical field at the end of a TEM waveguide}
\label{TEMWGfield}

This configuration has been studied in the literature \cite{T01}, but a self-contained derivation is provided here for completeness. We expand the optical electric field in frequency components $\Eb_\omega(\rb)$ using the notation \begin{align} \label{Enotation} 
\Eb(\rb,t)=\int\frac{d\omega}{2\pi}\,\Eb_\omega(\rb)\,\ee^{-\ii\omega t},
\end{align}
and take the incident electric field as a TEM wave with an electric field $\Eb_\omega^{\rm inc}(\rb)=\xx\,E^{\rm inc}_\omega\,\ee^{\ii kz}$ and wave vector $k=\omega/c$ in the $|x|<a/2$ region. The total field inside the waveguide can be expanded as
\begin{subequations}
\begin{align} \label{Ein} 
\Eb_\omega(\rb)=E^{\rm inc}_\omega\bigg\{\ee^{\ii kz}\,\xx+\sum_{n=0}^\infty r_n\,\ee^{\kappa_nz}\Big[\cos(q_nx)\,\xx+\frac{q_n}{\kappa_n}\sin(q_nx)\,\zz\Big]\bigg\} \quad\quad\quad\quad(z<0,\;|x|<a/2)
\end{align}
in terms of reflected waveguide modes with coefficients $r_n$ indexed by integers $n$. Here, we define $q_n=2\pi n/a$
and $\kappa_n=\sqrt{q_n^2-k^2-\ii0^+}$ with ${\rm Re}\{\kappa_n\}$>0 (in particular, $\kappa_0=-\ii k$). Note that only modes with even $E_x$ components are included, compatible with the symmetry of the incident field. The $n=0$ reflected mode is a propagating wave, whereas $n>0$ modes are evanescent but can contribute to the field near the aperture. From Faraday's law [$\Hb_\omega=(-\ii c/\omega)\nabla\times\Eb_\omega$], the corresponding magnetic field reduces to
\begin{align} \label{Hin} 
\Hb_\omega(\rb)=E^{\rm inc}_\omega\bigg[\ee^{\ii kz}+\sum_{n=0}^\infty r_n\,\frac{\ii k}{\kappa_n}\,\ee^{\kappa_nz}\cos(q_nx)\bigg]\,\yy  \quad\quad\quad\quad(z<0,\;|x|<a/2).
\end{align}
\end{subequations}
In the outer region ($z>0$), the electromagnetic field can be expressed as a combination of p-polarized waves with wave vector $q$ along $x$:
\begin{subequations}
\begin{align} 
&\Eb_\omega(\rb)=E^{\rm inc}_\omega\int\frac{dq}{2\pi}\,e_{q\omega}\,\ee^{\ii(qx+k_zz)}\,\Big(\xx-\frac{q}{k_z}\zz\Big) \quad\quad\quad\quad(z>0),  \label{Eout}\\
&\Hb_\omega(\rb)=E^{\rm inc}_\omega\int\frac{dq}{2\pi}\,e_{q\omega}\,\frac{k}{k_z}\,\ee^{\ii(qx+k_zz)}\,\yy \quad\quad\quad\quad\quad\quad\;\;\;(z>0),  \label{Hout}
\end{align}
\end{subequations}
where $k_z=\sqrt{k^2-q^2+\ii0^+}$ with ${\rm Im}\{k_z\}$>0.

Imposing the continuity of the $x$ component of the electric field at the $z=0$ plane, we can obtain the amplitudes $e_{q\omega}$ by applying the inverse Fourier transform to Eq.~(\ref{Eout}) with $E_{x,\omega}(x,y,0)$ given by Eq.~(\ref{Ein}) for $|x|<a/2$ and $E_{x,\omega}(x,y,0)=0$ for $|x|>a/2$ (i.e., the surface-parallel electric field vanishes at the metal). This leads to the expression
\begin{subequations}
\begin{align} 
e_{q\omega}=\frac{2\sin(qa/2)}{q}\bigg[1+\sum_{n=0}^\infty (-1)^n r_n\frac{q^2}{q^2-q_n^2}\bigg] \label{Eqalphan}
\end{align}
in terms of the coefficients $r_n$. Likewise, from the continuity of the in-plane magnetic field at $z=0$, performing the inverse discrete Fourier transform of Eq.~(\ref{Hin}) with $H_{y,\omega}(x,y,0)$ given by Eq.~(\ref{Hout}), and using the relation $\int_{-a/2}^{a/2}dx\,\cos(q_nx)\cos(q_{n'}x)=(a/2)\delta_{nn'}(1+\delta_{n0})$, we find
\begin{align} 
r_n=\delta_{n0}-\frac{2\ii(-1)^n}{1+\delta_{n0}}\frac{\kappa_n}{\pi a}\int dq\,\frac{q}{k_z}\frac{\sin(qa/2)}{q^2-q_n^2}\,e_{q\omega}. \label{alphanEq}
\end{align}
\end{subequations}
Finally, inserting Eq.~(\ref{Eqalphan}) into Eq.~(\ref{alphanEq}) and truncating the sum to $n\le N$, we find the solution
\begin{subequations}
\label{Esolution}
\begin{align} 
r=M^{-1}\cdot b,
\end{align}
where $r$ and $b$ are vectors of $N+1$ components given by $r_n$ (with $n=0,\cdots,n$) and
\begin{align} 
b_n=\delta_{n0}-I_{n0},
\end{align}
respectively, while $M$ is an $(N+1)\times(N+1)$ matrix of components
\begin{align} 
M_{nn'}=\delta_{nn'}+I_{nn'}.
\end{align}
Here, we have defined the integrals
\begin{align} 
I_{nn'}=\frac{4\ii(-1)^{n+n'}}{1+\delta_{n0}}\frac{\kappa_{n}}{\pi a}\int dq\,\frac{q^2}{k_z}\frac{\sin^2(qa/2)}{(q^2-q_n^2)(q^2-q_{n'}^2)}.
\end{align}
\end{subequations}
We solve Eqs.~(\ref{Esolution}) numerically and find converged results with $N=0$ when the light wavelength $\lambda=2\pi c/\omega$ exceeds a few times the slit width $a$ (see Fig.~\ref{FigS4}). This is the case, for example, of the frequency-dependent reflection coefficient, which is given by $r_0$ [i.e., the coefficient of the downward-propagating TEM mode $n=0$ in Eq.~(\ref{Ein})]. This quantity is plotted in Fig.~\ref{FigS4}a as a function of the wavelength-to-width ratio $\lambda/a$.

\begin{figure*}[htb]
\centering \includegraphics[width=1.00\textwidth]{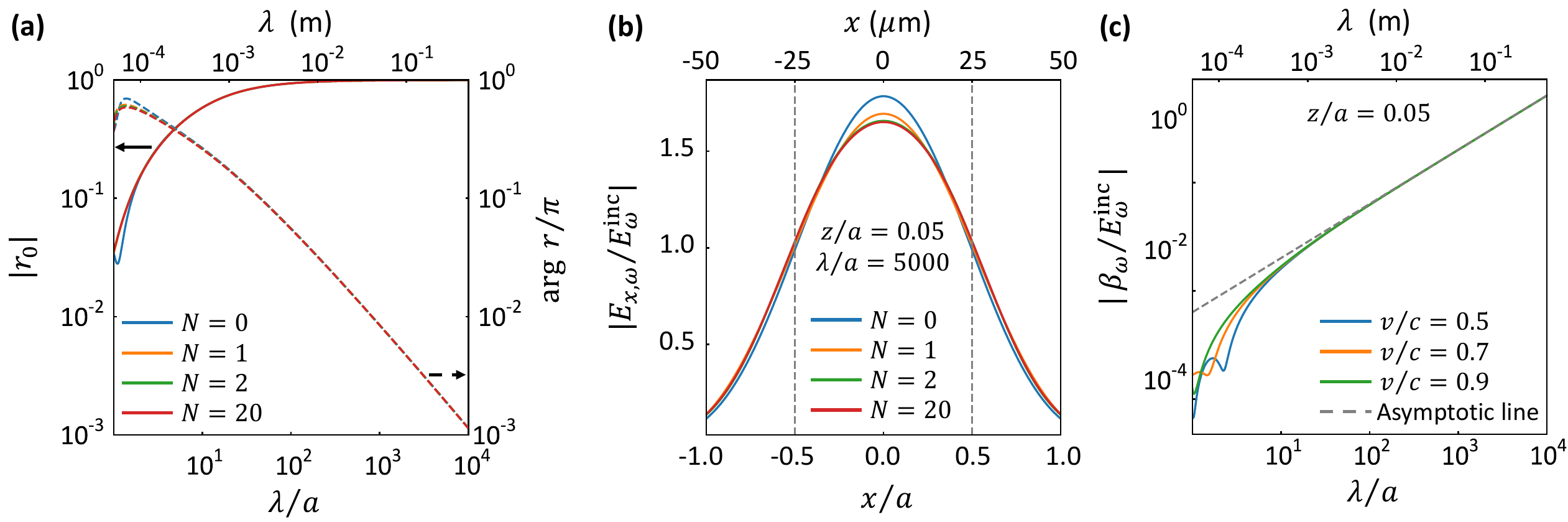}
\caption{{\bf Numerical results for the TEM waveguide in Fig.~\ref{FigS3}.}
{\bf (a)}~Reflection coefficient $r$ calculated from Eqs.~(\ref{Esolution}) for different values of the maximum reflected-mode order $N$ as a function of the wavelength-to-width ratio $\lambda/a=2\pi c/a\omega$.
{\bf (b)}~Electric field component along $x$ evaluated at $z=0.05\,a$ for $\lambda/a=5000$.
{\bf (c)}~Electron--light coupling coefficient for an electron passing at a distance $z=0.05\,a$ from the waveguide [Eq.~(\ref{betawcalc})]. We consider different electron velocities and compare the results to the $\lambda\gg a$ asymptotic limit [dashed line, Eq.~(\ref{betawasymp})].
For optical wavelengths exceeding a few times the slit width, the $N=0$ calculations (only the $n=0$ reflected mode is included) yield an accurate description of $r$, the near field $E_{x,\omega}$, and the coupling coefficient $\beta_\omega$. The upper horizontal scale corresponds to $a=50$~nm.}
\label{FigS4}
\end{figure*}

\subsection{Electron--light coupling coefficient}
\label{WGbetaw}

Once the coefficients $r_n$ are numerically calculated from Eqs.~(\ref{Esolution}), the $x$ component of the electric field $E_{x,\omega}$ in the $z>0$ region is obtained by plugging Eq.~(\ref{Eqalphan}) into Eq.~(\ref{Eout}), leading to the density plot shown in Fig.~\ref{Fig4}a of the main text (see also Fig.~\ref{FigS4}b). The electron--light coupling coefficient is then obtained by inserting this field into Eq.~(\ref{betaw}), with $x$ substituted for $z$ because the electron velocity is parallel to $\xx$ in the present configuration. We find
\begin{align} \label{betawcalc} 
\beta_\omega=\frac{2e}{\hbar\omega}\;E^{\rm inc}_\omega\;\ee^{-\omega z/v\gamma}
\;\frac{\sin(\omega a/2v)}{(\omega/v)}\bigg[1+\sum_{n=0}^\infty (-1)^n r_n\frac{\omega^2}{\omega^2-q_n^2v^2}\bigg].
\end{align}
For the parameters considered in this work (electron velocities amounting to a substantial fraction of the speed of light and optical wavelengths largely exceeding both the slit width and the electron--waveguide separation), we can neglect $n>0$ terms in Eq.~(\ref{betawcalc}), approximate $r_0\approx1$ (see Fig.~\ref{FigS4}a), and consider $a,z\ll v/\omega$. Then, the coupling coefficient reduces to
\begin{align} \label{betawasymp} 
\beta_\omega\approx\frac{2ea}{\hbar\omega}\;E^{\rm inc}_\omega.
\end{align}
We find that this expression yields accurate results under the conditions considered in this work (see dashed line in Fig.~\ref{FigS4}c).

\subsection{Optical energy}
\label{opticalenergy}

From the Poynting vector of the incident TEM wave, the externally supplied optical energy reduces to
\begin{align} \nonumber 
\mathcal{E}_{\rm opt}=\frac{c\,aL_y}{8\pi^2}\int d\omega \,\big|E^{\rm inc}_\omega\big|^2
=\frac{c\,aL_y}{4\pi^2}\int_0^\infty d\omega \,\big|E^{\rm inc}_\omega\big|^2,
\end{align}
where $L_y$ is the waveguide length along the slit direction $y$ (see Fig.~\ref{FigS3}). Replacing $E^{\rm inc}_\omega$ by $\beta_\omega$ according to Eq.~(\ref{betawasymp}) and comparing the result to Eq.~(\ref{Fn1}), we find that the TEM waveguide configuration corresponds to $n=1$ with
\begin{align} \label{CTEM} 
C=\frac{\hbar}{32\pi^2\alpha}\frac{L_y}{a},
\end{align}
where $\alpha=e^2/\hbar c\approx1/137$ is the fine-structure constant. Combining the TEM waveguide configuration with the optimization scheme ({\it i}) (Sec.~\ref{scheme1}), we obtain 
\begin{align} 
\mathcal{E}_{\rm opt}
&=\frac{3\hbar}{64\alpha\,\eta_0^2\,\sigmat_i^4}\;\frac{1}{(\omega_{\rm  max}^3-\omega_{\rm min}^3)}\frac{L_y}{a} \nonumber
\end{align}
by using Eqs.~(\ref{Fn2}), (\ref{scheme1Fn}), and (\ref{CTEM}).

\begin{figure*}[htb]
\centering \includegraphics[width=0.5\textwidth]{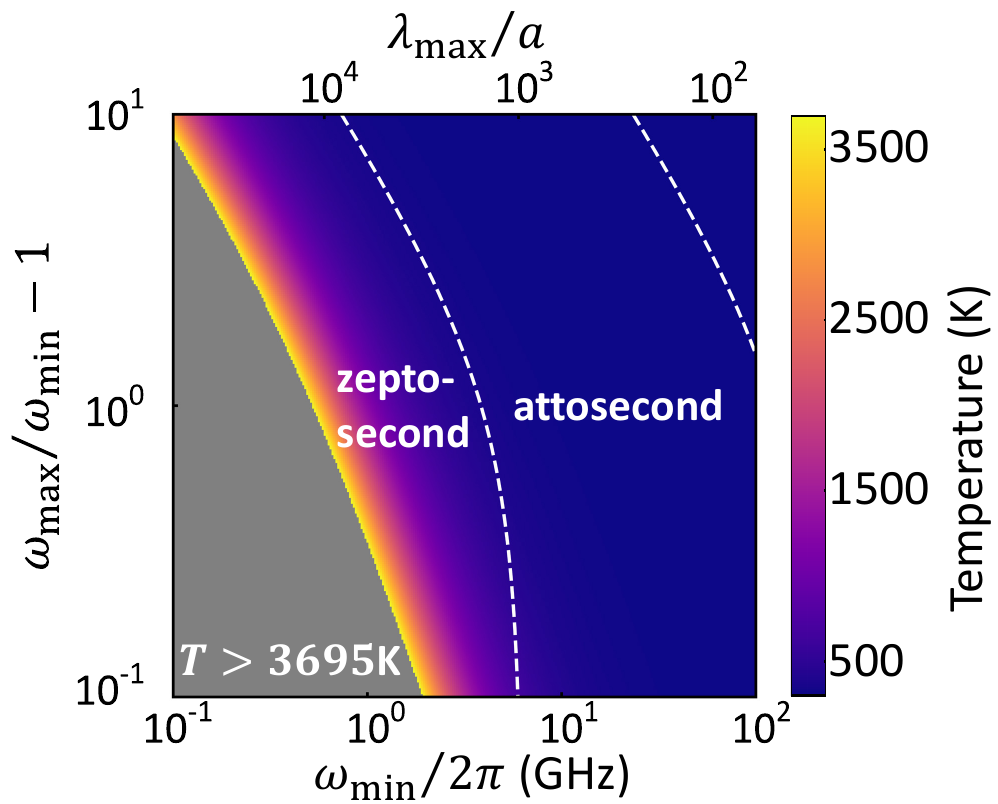}
\caption{{\bf Optical heating of the metal walls in the waveguide.} We show a density plot with an upper estimate of the temperature reached at the metal edge upon pulse irradiation under the same conditions as in Fig.~\ref{Fig4}c of the main text. The scale is saturated at the melting temperature of W (3695~K). The same zeptosecond and attosecond compression regions as in Fig.~4c are indicated here for reference.}
\label{FigS5}
\end{figure*}

\subsection{Optical heating}

A possible problem in the present configuration relates to the potential damage produced in the metal by Joule heating from the optical wave. We estimate the temperature elevation in the material by taking the fields calculated inside the waveguide opening and matching them to the field within the metal, assuming a large finite conductivity $\sigma$ \cite{J99}. In the $\lambda\gg a$ limit, setting the reflectivity to $r_0=1$ and neglecting evanescent waves (i.e., taking $N=0$, which provides a reasonable description in this limit, as shown in Fig.~\ref{FigS4}), the field reduces to [see Eq.~(\ref{Hin}) as well as Eq.~(8.10) of Ref.~\cite{J99}, written here in CGS units]
\begin{align} \label{Ersd} 
\Eb_\omega^{\rm metal}(\rb)\approx-{\rm sign}\{x\}\,\zz\,\sqrt{\frac{\omega}{\pi\sigma}}\,\ee^{\ii\pi/4}\,E^{\rm inc}_\omega\,\ee^{(\ii-1)(|x|-a/2)/\delta_\omega} \sin(kz)
\end{align}
for $|x|>a/2$ (i.e., in the metal region close to the waveguide walls), where $\delta_\omega=c/\sqrt{2\pi\omega\sigma}$ is the frequency-dependent skin depth. We now use the Drude model to calculate the energy absorbed per unit volume inside the metal as $d\mathcal{E}_{\rm abs}/d\rb=(\sigma/2\pi)\int d\omega\,|\Eb_\omega(\rb)|^2$, which, combined with Eq.~(\ref{Ersd}) and averaged along the direction $z$ parallel to the waveguide, becomes
\begin{align} \label{dEdr} 
\frac{d\mathcal{E}_{\rm abs}}{d\rb}=\frac{1}{4\pi^2}\int d\omega\;\omega\;\big|E^{\rm inc}_\omega\big|^2\,\ee^{-(2|x|-a)/\delta_\omega}.
\end{align}
Now, if we adopt scheme ({\it i}) (Sec.~\ref{scheme1}), set $n=1$ (Sec.~\ref{opticalenergy}), and take $a/\lambda\ll1$ (Sec.~\ref{WGbetaw}) then $E^{\rm inc}_\omega$ is obtained by combining Eqs.~(\ref{betawfw}), (\ref{fwtscheme1}), and (\ref{betawasymp}), which lead to
\begin{align} \label{Ewinc} 
E^{\rm inc}_\omega=\frac{-3\ii\pi\hbar\omega}{4\,\eta_0\,e\,\sigma_i^2\,a\,(\omega_{\rm max}^3-\omega_{\rm min}^3)}.
\end{align}
As expected, maximum heating occurs at the metal walls ($x=\pm a/2$) according to Eq.~(\ref{dEdr}), which, upon insertion of Eq.~(\ref{Ewinc}), reduces to a metal-independent result (in the large conductivity limit):
\begin{align} \nonumber 
\frac{d\mathcal{E}_{\rm abs}}{d\rb}=\frac{9\,\hbar^2}{128\,\eta_0^2\,e^2\,\omega_{\rm min}^2\,\sigma_i^4\,a^2}\;\frac{r^4-1}{(r^3-1)^2}
\end{align}
with $r=\wmax/\wmin$. Energy is initially absorbed by the conduction electrons. However, damage to the material is mainly produced when the rise in atomic-lattice temperature exceeds a certain material-dependent threshold. As an upper bound to the maximum temperature $T$ reached after propagation of a GHz wave, we neglect heat diffusion and obtain the equilibrium temperature $T$ from the condition that the sum of electronic heat ($\gamma T^2/2$, where $\gamma$ is a metal-dependent coefficient \cite{K1987}) and lattice heat ($c_lT$, where $c_l$ is the lattice heat capacity) densities is equal to $d\mathcal{E}_{\rm abs}/d\rb$ plus the thermal energy density at room temperature $T_0$ before irradiation (i.e., $\gamma T_0^2/2+c_lT_0$). For tungsten ($\gamma\approx106\,$J/m$^3$K$^2$ \cite{S1983_2} and $c_l=2.55\times10^6\,$J/m$^3$K$^{-1}$ \cite{L05}), the upper estimate lies below the melting temperature (3695~K) within a wide region in which zeptosecond compression is achieved. In a real system, electronic heat diffusion should lead to substantially lower temperatures during the short microsecond duration of the pulse.

\end{widetext}

\end{document}